\documentclass[aps,pre,amsmath,amssymb,lengthcheck,showpacs,superscriptaddress]{revtex4-1}
\usepackage{graphicx}
\usepackage{color}
\usepackage{amsfonts}
\usepackage{physics}
\usepackage{braket}
\usepackage{multibib}
\usepackage[english]{babel}

\begin{document}


\title{Phase separation kinetics of a symmetric binary mixture of glass-forming liquids}

\author{Yuri Oku}
\affiliation{Department of Earth and Space Science, Osaka University, Toyonaka, Osaka 560-0043, Japan}
\author{Kyohei Takae}
\affiliation{Department of Fundamental Engineering, Institute of Industrial Science, University of Tokyo, Tokyo 153-8505, Japan}
\author{Atsushi Ikeda}
\affiliation{Graduate School of Arts and Sciences, University of Tokyo, Tokyo 153-8902, Japan}
\affiliation{Research Center for Complex Systems Biology, Universal Biology Institute, University of Tokyo, Komaba, Tokyo 153-8902, Japan}

\date{\today}


\begin{abstract}
Mixtures of glass-forming fluids sometimes exhibit glass-glass phase separation at low temperatures. 
Here, we use a molecular dynamics simulation to study one of the simplest examples of the glass-glass phase separation. 
We consider a mixture composed of type A and B particles, in which the A-A and B-B interactions are the identical Lennard-Jones interactions and the A-B interaction is repulsive only. 
To avoid crystallization, we also introduce the polydispersity in the particle sizes for each component. 
We study the phase separation kinetics of this model at a 50:50 concentration at various temperatures. 
We find that hydrodynamic coarsening takes place when the temperature is higher than the onset temperature of the glassy dynamics. 
At lower temperatures, diffusive coarsening is observed over a long duration, and a further slower coarsening appears within a shorter time. 
Below the glass transition temperature, the domain growth does not stop but becomes logarithmically slow or even slower than logarithmic. 
By analyzing two-time correlation functions, we show that these slow coarsening processes are accompanied by a slowing down of the microscopic dynamics, which has qualitative similarities with the aging dynamics without phase separation. 
Based on the results, we discuss a possible link between the slow coarsening and the aging-like microscopic slowing down in the glass-glass phase separation. 
\end{abstract}


\maketitle

\section{Introduction}

Fluid mixtures, alloys, various soft matters, and biological matters frequently exhibit immiscibility~\cite{Furukawa1985,1992solids,Bray1994,onuki2007phase,Khachaturyan2008,Tanaka2012a,Berry2018}. 
When these systems are brought into the immiscible region in a phase diagram, they undergo phase separation and exhibit inhomogeneous spatial patterns of components. 
A wide variety of spatial patterns and their evolution kinetics have been found and studied.
We here focus on the phase separation of a nearly equal concentration mixture of two types of particles, such as binary alloys and binary fluids. 
In both alloys and fluids, homogeneous states become unstable in the early stage of phase separation, and interpenetrating domains that are separated by domain walls appear~\cite{Cahn1961,Langer1975,Kawasaki1978}. 
These domains slowly coarsen in the late stage of phase separation. 
In binary alloys, hydrodynamic transport is absent and the domain growth is driven by the diffusion of particles. 
The characteristic size of the domains grows as $\xi \propto t^{1/3}$~\cite{Lifshitz1961}, which is referred to as diffusive coarsening. 
On the other hand, in binary fluids, the transport of particles by hydrodynamic flow significantly accelerates the domain growth. 
In particular, the tube instability results in $\xi \propto t$ in three-dimensional $d=3$ fluids~\cite{Siggia1979}. 
In $d=2$ fluids, the tube instability is not operative~\cite{SanMiguel1985,Furukawa1985}, but inertial effects lead to $\xi \propto t^{2/3}$~\cite{Furukawa1985}.
These coarsening processes are referred to as hydrodynamic coarsening. 
Diffusive and hydrodynamic coarsening have been established via numerical simulations of mesoscopic models~\cite{Oono1987,Farrell1989,Koga1993,Osborn1995,Wagner1998,Gonzalez-Segredo2003}, molecular dynamics (MD) simulations~\cite{Velasco1993,Velasco1996,Ahmad2012}, experiments~\cite{Hennion1982,Chou1981,Wong1978,Goldburg1978,Takenaka1992}, and theoretical arguments based on the Cahn-Hilliard equation and hydrodynamic equations~\cite{Bray1994,1992solids,Furukawa1985,onuki2007phase}.

Fluids typically undergo a glass transition at low temperatures if quenched sufficiently rapidly~\cite{Debenedetti1996}. 
In fluid mixtures, the glass phase is sometimes located in the immiscible region in a phase diagram. 
How does the phase separation proceed in these systems?
Because the glass transition dramatically changes the dynamics of the system~\cite{Berthier2011b}, it also affects the phase separation kinetics. 
Experimentally, phase separation in glasses has been observed in multicomponent oxide glasses~\cite{Mazurin1984,Doremus1994} and metallic glasses~\cite{Kim2013a}. 
It is known that after a rapid quench into the glass phase, the phase separation virtually stops.
The characteristic size of the domains typically remains submicroscopic, depending on the thermal history~\cite{Doremus1994,Kim2013a}. 
The phase separation can be restarted by thermal annealing. 
The morphology of the domains and the domain growth kinetics in this process have been studied using light scattering, atomic force microscopy, X-ray tomography, and so on~\cite{ELMER1970,Simmons1974,Malik1998,Dalmas2007,Wheaton2007,Bouttes2014,Bouttes2015}.  
Diffusive coarsening was reported for the case where the annealing temperature is not far from the glass transition temperature~\cite{Malik1998,Dalmas2007,Wheaton2007}, while hydrodynamic coarsening was reported for the case in which the annealing temperature is much higher than the glass transition temperature~\cite{Bouttes2014,Bouttes2015}. 
From the application point of view, phase separation offers a chance to control the material properties of glasses and create functional glasses~\cite{Doremus1994}. 

To obtain a clearer picture of the phase separation in the glass phase, numerical simulations can be an ideal tool because both the microscopic and mesoscopic dynamics of a system can be studied under well-controlled conditions. 
Indeed, the phase separation of model glass formers is often observed in MD simulations~\cite{Ninarello2017}. 
However, the domain structures and domain growth kinetics are rarely studied quantitatively because many studies are interested primarily in the glass transition dynamics itself. 
Exceptionally, there are several numerical studies on phase separation into components whose glass transition temperatures are widely different, such that one of the components forms a glass while the others remain fluid~\cite{Sappelt1997,Furukawa2010,Testard2011,Testard2014,Chaudhuri2016,Tateno2019,Douglass2019}.
One interesting point in this case is the presence of large dynamic asymmetry between the components, which can lead to unusual domain structures and growth kinetics~\cite{Tanaka2012a}. 
This case has also attracted attention in the context of colloidal gelation~\cite{Zaccarelli2007,Lu2008}. 
Refs.~\cite{Testard2011,Testard2014,Chaudhuri2016} considered the simplest example of this case. 
They studied the quench dynamics of Lennard-Jones particles at low temperatures at various densities using MD simulations. 
At the densities where the gas-liquid phase separation takes place at intermediate temperatures, the gas-glass phase separation appears at lower temperatures. 
They found that the domain growth becomes logarithmically slow in this regime and that intermittent glassy dynamics drives the very slow phase separation~\cite{Testard2014}. 
On the other hand, Refs.~\cite{Furukawa2010,Tateno2019} took into account the hydrodynamic interaction among particles and revealed its impact on the phase separation in colloidal suspensions and the gelation. 

In this work, we study the simplest example of glass-glass phase separation. 
We consider a Lennard-Jones mixture composed of type A and B particles, in which the A-A and B-B interactions are identical and the A-B interaction is different. 
To avoid crystallization, we also introduce the polydispersity in the particle sizes for each component. 
Because type A and B particles are identical, there is no dynamic asymmetry in this model; the system will exhibit a phase separation into A-rich glass and B-rich glass at low temperatures.  
To the best of our knowledge, this case has not been studied in detail by numerical simulations. 
We use an MD simulation to study the phase separation kinetics of this model at a 50:50 concentration at various temperatures. 
We find that hydrodynamic coarsening takes place when the temperature is higher than the onset of the glassy dynamics. 
At lower temperatures, diffusive coarsening is observed over a long duration, and further slower coarsening appears after a shorter time. 
Below the glass transition temperature, the domain growth does not stop but becomes logarithmically slow or even slower than logarithmic. 
By analyzing the microscopic dynamics during phase separation, we show that this slow coarsening is accompanied by a slowing down of the dynamics at the microscopic level. 
This slowing down has qualitative similarities with the aging dynamics without phase separation. 
However, quantitatively, the relaxation during the phase separation is faster than that without phase separation, suggesting that the inhomogeneous concentration field accelerates the microscopic relaxation in the former case. 
Finally, we discuss a possible mechanism of the slow coarsening process by a simple extension of the Cahn-Hilliard equation to take into account the aging-like microscopic slowing down. 

This article is organized as follows. In Sec. II, we define the model and describe the numerical simulations. 
In Sec. III, we calculate the static phase diagram of the model. 
In Sec. IV, we study the glassy dynamics in the one-component version of the model. 
In Sec. V, we study the kinetics of the domain growth during phase separation. 
In Sec. VI, we study the microscopic dynamics during phase separation and discuss the impact of its slowing down on the phase separation kinetics. 
In Sec. VII, we summarize our results.

\section{Model and Method}

\subsection{Model}
We consider a binary mixture of particles (type A and B) for $d=2$. 
Particles of the same species interact through the Lennard-Jones (LJ) potential: 
\begin{equation}
u_{AA}(r) = u_{BB}(r) = 4 \epsilon \left[\left(\frac{a_{ij}}{r}\right)^{12} - \left(\frac{a_{ij}}{r}\right)^{6} \right], 
\end{equation}
where $r$ is the distance between particles, $a_{ij} = (a_i + a_j)/2$ and $a_i$ is the diameter of particle $i$. 
Particles of different species interact through the Weeks-Chandler-Andersen (WCA) potential: 
\begin{equation}
u_{AB}(r) = 
\begin{cases}
4 \epsilon \left[\left(\frac{a_{ij}}{r}\right)^{12} - \left(\frac{a_{ij}}{r}\right)^{6} \right] + \epsilon, & r<2^{1/6} a_{ij}, \\
0, & r \ge 2^{1/6} a_{ij}. 
\end{cases}
\end{equation}
Thus, in this model, particles have attractive interactions with other particles of the same species and repulsive interactions with those of different species. 

On top of this bidispersity, we introduce the polydispersity of the particle diameters to avoid crystallization. 
The diameters of particles are distributed uniformly between $0.8a$ and $1.2a$ for each component. 
The polydispersity index, i.e., the standard deviation of the particle diameter, is $\delta \approx 11.5 \%$. 
All particles have the same mass $m$.  
The particles are put into a two-dimensional cell of area $V$ with the periodic boundary condition. 
The number density $\rho = N/V = 0.925 a^{-2}$. 
All the results are reported in Lennard-Jones units: length in $a$, energy in $\epsilon$, temperature in $\epsilon/k_B$, and time in $\sqrt{m a^2/\epsilon}$. 
In practice, we truncate and shift the LJ potential at the cutoff length of $3 a$. 

The numbers of A and B particles are denoted as $N_A$ and $N_B$, with $N_A + N_B = N$, and the fraction of A particles is defined as $x = N_A/N$. 
The total number of particles $N=20000$ unless otherwise noted. 
We mainly consider the model at $x = 0.5$ and refer to this case as the ``binary model''. 
We also refer to the case $x = 1$ as the ``pure model''. 
We note that the mechanical and vibrational properties of the glass state of the pure model were studied in Ref.~\cite{Tanguy2002}. 

\subsection{MD simulations}

We perform constant-temperature MD simulations of the pure and binary models to study their dynamics~\cite{allen1989computer}. 
We numerically integrate the Newtonian equations of motion by the velocity Verlet algorithm with the time step of $10^{-2}$. 
The temperature is controlled by the velocity rescaling method.  
To study the phase separation kinetics in the binary model and the aging dynamics in the pure model, we first perform MD simulations at $T=100$ to obtain the equilibrated configurations at this temperature.
Then, starting from these configurations, we perform MD simulations at several target temperatures. 
This protocol corresponds to the instantaneous quench from $T=100$ to the target temperatures.  
The target temperatures are $T=$ 0.8, 0.5, 0.4, 0.3, 0.2, 0.1, and 0.03. 
For each temperature, three independent simulations are performed (six for $T=0.4$), and the final results are obtained by averaging the results of the independent runs. 
To study the equilibrium dynamics in the pure model, we first perform MD simulations at $T=0.8$ to equilibrate the system at this temperature. 
Starting from the obtained configuration, we gradually decrease the temperature and equilibrate the system at each target temperature. 
The simulation time of the equilibration runs is fixed at $10^5$. 
Then, we perform the production runs at each target temperature. 

\subsection{MC simulations}

We also perform Monte Carlo (MC) simulations of the binary model~\cite{frenkel2001understanding}. 
As in the MD simulations, we equilibrate the system at $T=100$ and suddenly quench the system to the target temperatures. 
The MC simulations are performed using a simple displacement move with the Metropolis rule. 
Each MC cycle consists of $N$ trial displacements in which we attempt to displace particles over a square, the dimensions of which were chosen such that the acceptance ratio becomes approximately 50 \% for each temperature. 

To determine the static phase diagram, we also perform semi-grand canonical MC simulations~\cite{frenkel2001understanding,Miguel1995}. 
In these simulations, we focus on the size-monodisperse version of the present model for simplicity. 
We also set $N=4000$ for these simulations. 
The semi-grand canonical MC simulation is used to generate a series of configurations at a fixed temperature, with equal chemical potentials of the A and B particles. 
Each MC cycle consists of $N$ trial swaps of the particle type in addition to $N$ trial displacements. 
In each trial swap, we choose one particle in a box randomly and attempt to change its type from A to B or from B to A. 
We perform this simulation up to $10^6$ MC steps at temperatures $1.6 \leq T \leq 2.5$. 

\section{Equilibrium phase diagram}

We first determined the static phase diagram of our model. 
We focused on the size-monodisperse version of the model for simplicity and performed semi-grand canonical MC simulations to calculate the coexistence temperatures of two liquid phases. 
For the simulation details, see Sec.~II~C. 

\begin{figure}[t]
\centering
\includegraphics[width=\columnwidth]{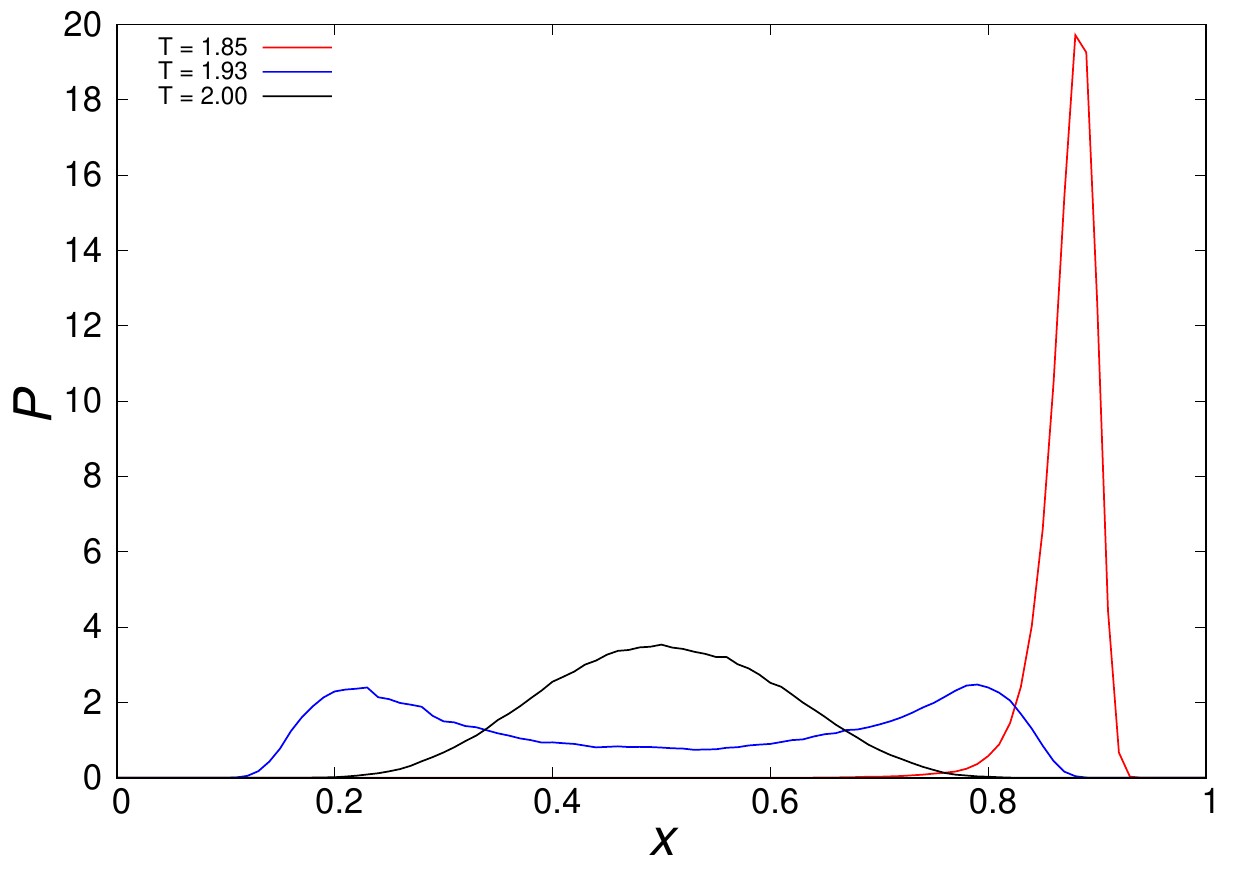}
\caption{
The probability distribution of the fraction of A particles $P(x,T)$ obtained by the semi-grand canonical MC simulations.
}
\label{fig1}
\end{figure}

\begin{figure}
\centering
\includegraphics[width=\columnwidth]{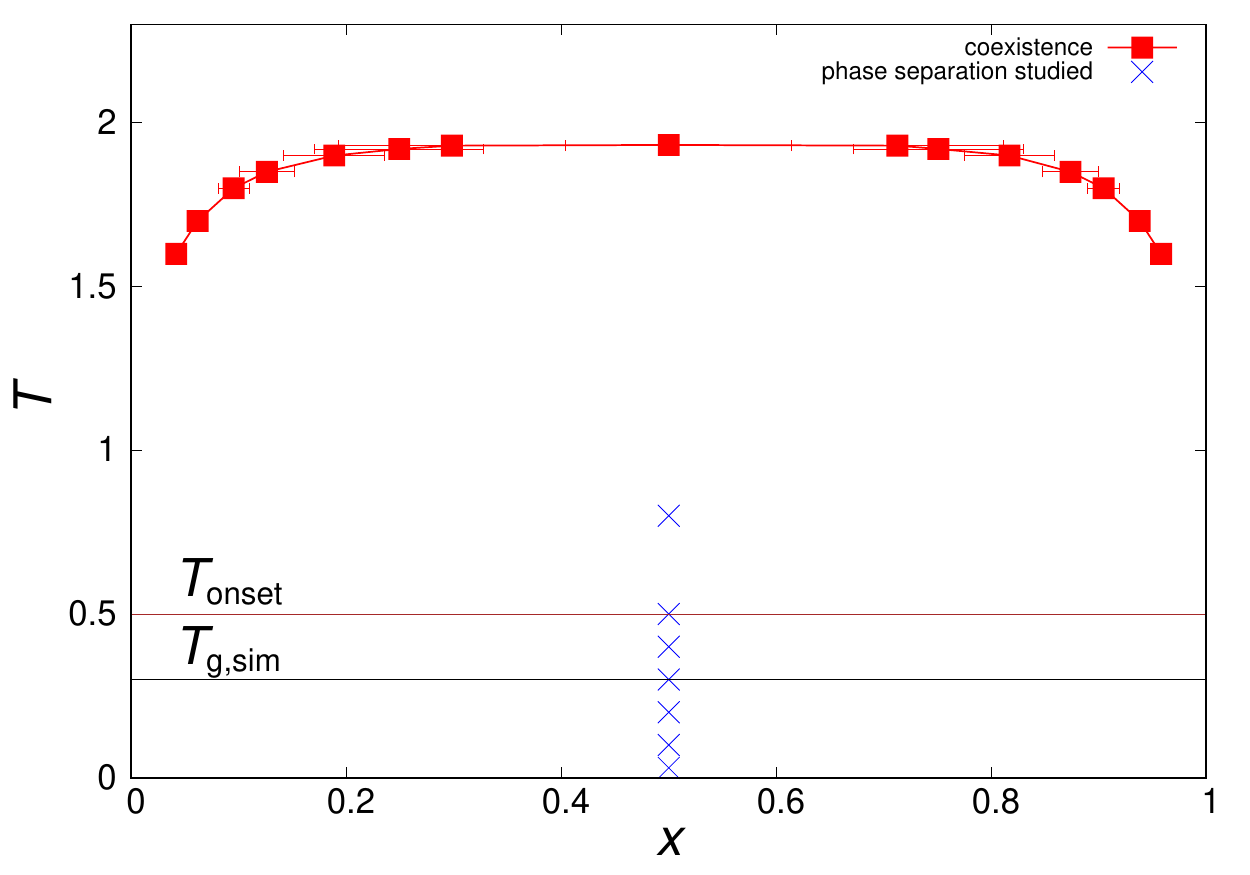}
\caption{
Phase diagram of the model.
Closed squares show the coexistence temperatures $T_{\rm coex}(x)$ for the liquid-liquid equilibria of the size-monodisperse system. 
The estimate for the critical point ($T_c = 1.932$ at $x = 0.5$) is included. 
The temperatures at which the phase separation kinetics are studied are indicated by the blue crosses. 
The onset temperature of the glassy dynamics $T_{\rm onset} = 0.5$ and the glass transition temperature $T_{\rm g,sim} = 0.3$ of the pure model are indicated by the thin solid lines. 
}
\label{fig2}
\end{figure}

In each semi-grand canonical MC simulation, the fractions of components fluctuate with the MC steps. 
Using the generated configurations, we calculated $P(x,T)$, the probability distribution of the fraction of A particles $x$ at temperature $T$. 
Fig.~\ref{fig1} shows the results at three temperatures. 
At high temperature $T \geq 1.95$ (see $T=2.00$ in Fig.~\ref{fig1}), $P(x,T)$ is unimodal with a peak at $x = 0.5$. 
This result means that the A and B particles mix well without phase separation. 
At low temperature $T \leq 1.85$ (see $T=1.85$ in Fig.~\ref{fig1}), $P(x,T)$ is unimodal with a peak at $x > 0.5$ or $x < 0.5$, meaning that the equilibrium configurations of A-rich or B-rich phases on the coexistence line are generated by the simulations. 
We calculated the coexistence concentrations as the average $\overline{x}(T) = \int x P(x,T) dx$ and obtained the coexistence temperatures $T_{\rm coex}(x)$ as the inverse function of $\overline{x}(T)$. 
We then symmetrized $T_{\rm coex}(x)$ about $x=0.5$ because the coexistence curve is symmetric in our model by definition. 
At intermediate temperatures (see $T=1.93$ in Fig.~\ref{fig1}), we obtained the bimodal distribution with peaks at $x > 0.5$ and $x < 0.5$. 
In this case, we fitted $P(x,T)$ into two asymmetric Gaussian functions and calculated $\overline{x}(T)$ for the A-rich and B-rich phases separately~\cite{Miguel1995}. 
In Fig.~\ref{fig2}, we plot the obtained coexistence temperatures, with the error bars estimated by the standard deviation of the concentrations. 
Because the critical behavior of the demixing of binary mixtures is expected to be in the Ising universality class, we fitted $T_{\rm coex}(x)$ into the power law $|x - 0.5| \propto (T_c - T)^{\beta}$ with $\beta = 1/8$. 
We found that our numerical data are consistent with this power law, and we obtained the critical temperature $T_c = 1.932$, which is also included in Fig.~\ref{fig2}. 

Although the phase diagram obtained here is for the size-monodisperse version of the model, we consider it to be a good reference for the original size-polydisperse systems. 
This is because the impact of the polydispersity is known to be perturbative only when the polydispersity is weak~\cite{Sollich2001}. 
In particular, the change in the coexistence temperatures due to the polydispersity is known to be proportional to the variance in the size distribution $\delta^2$, which is small in the present model~\footnote{The calculation of the prefactor to $\delta^2$ requires knowledge of the three-body correlation function. Unfortunately, this is computationally demanding; hence, we did not try to evaluate it.}. 

\section{Equilibrium dynamics in the pure model}

Before studying the phase separation kinetics of the binary model ($x=0.5$), we study the equilibrium dynamics of the pure model ($x=1$) for future reference. 
Because the pure model involves standard polydisperse LJ particles, it will exhibit glassy dynamics at low temperatures without phase separation. 
We performed MD simulations of the pure model at several target temperatures. 
For the simulation details, see Sec.~II~B. 

\begin{figure}[t]
\centering
\includegraphics[width=\columnwidth]{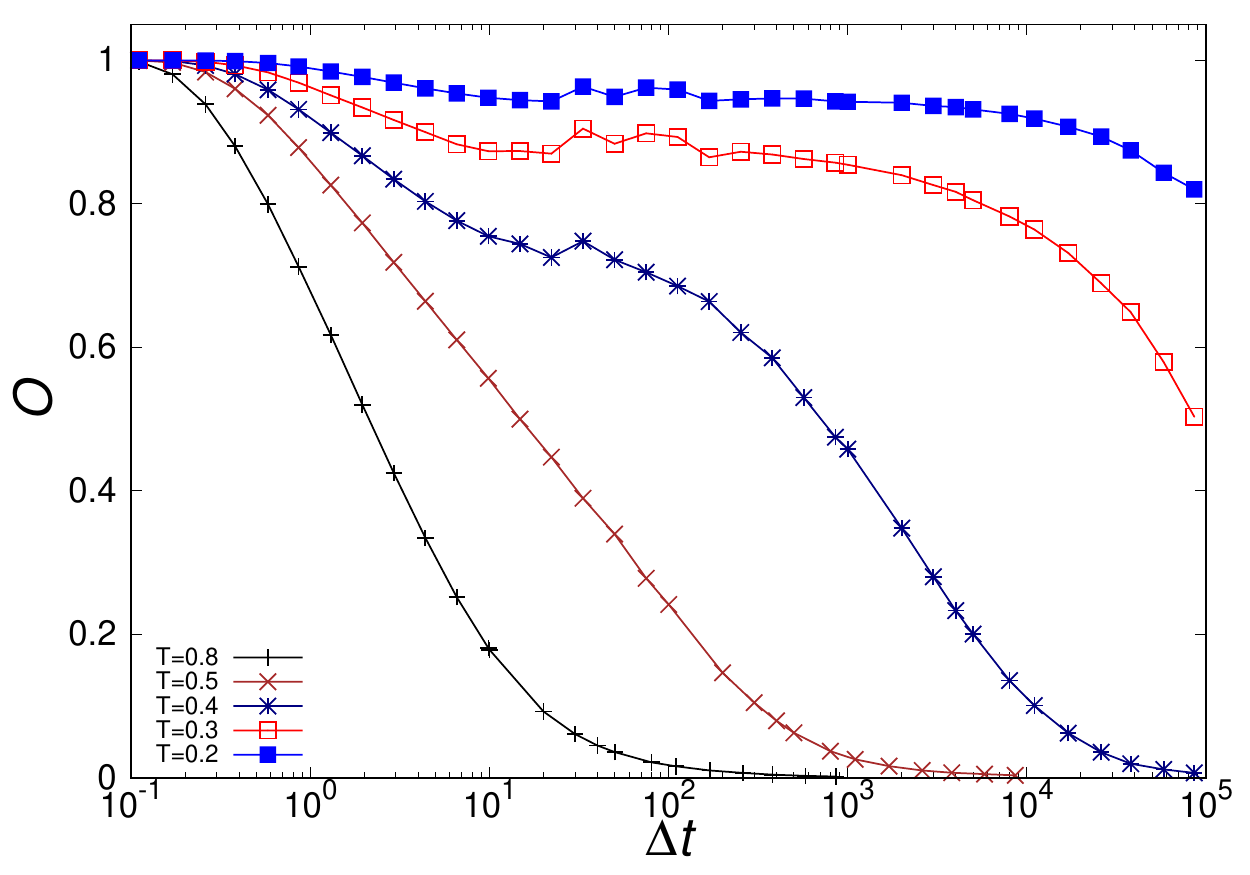}
\caption{
The self-part of the overlap function $O(\Delta t)$ of the pure model ($x = 1$).  
The temperatures are $T =$ 0.8, 0.5, 0.4, 0.3, and 0.2, from left to right. 
}
\label{fig3}
\end{figure}

To characterize the dynamics, we calculate the self-part of the overlap function
\begin{eqnarray}
O(\Delta t) = \frac{1}{N} \sum_{i} \langle \theta (|\vec{r}_i(\Delta t) - \vec{r}_i(0)| - \ell) \rangle, 
\end{eqnarray}
where $\vec{r}_i(t)$ is the position of $i$-th particle at time $t$, $\langle \cdot \rangle$ denotes the time translation average and we set $\ell = 0.3$ to monitor the microscopic rearrangements of particles. 
Fig.~\ref{fig3} shows the results. 
The pure model exhibits the canonical glassy behavior at low temperatures~\cite{Berthier2011b}. 
Normal rapid relaxation is observed at $T=0.8$. 
Stretched relaxation sets in at $T = 0.5$; thus, we define the onset temperature of the glassy dynamics as $T_{\rm onset} = 0.5$. 
As the temperature is further decreased, the two-step relaxation becomes apparent, and the relaxation becomes drastically slower. 
At $T \leq 0.3$, the overlap function does not reach zero in our simulation time, namely, the model virtually vitrifies in our simulation time. 
We define the simulation glass transition temperature as $T_{\rm g,sim} = 0.3$. 
These two characteristic temperatures are indicated as lines in Fig.~\ref{fig2}. 

The dynamical quantities of two-dimensional glass formers are known to be influenced by the Mermin-Wagner fluctuations~\cite{Shiba2012,Flenner2015,Shiba2016}. 
Because this effct accelerates the decay of the overlap function and blurs the intrinsic slow relaxation, it can be a cause of the underestimation of the onset and glass transition temperatures. 
Because this effect is greater in larger system, the onset and glass transition temperatures could be better estimated from simulations on smaller systems. 
Therefore, we also performed the MD simulations of $N=1000$ system.
We confirmed that, although $N=1000$ system indeed exhibits slightly slower relaxation, $T_{\rm onset} = 0.5$ and $T_{\rm g,sim} = 0.3$ are still reasonable even for $N=1000$ system. 
Therefore, our later discussion based on these temperatures are not affected much by the Mermin-Wagner fluctuations. 

\section{Phase separation kinetics in the binary model}

\subsection{Qualitative observations}

\begin{figure*}[t]
\begin{tabular}{llll}
\begin{minipage}{0.25\hsize}
\includegraphics[width=1.02\columnwidth]{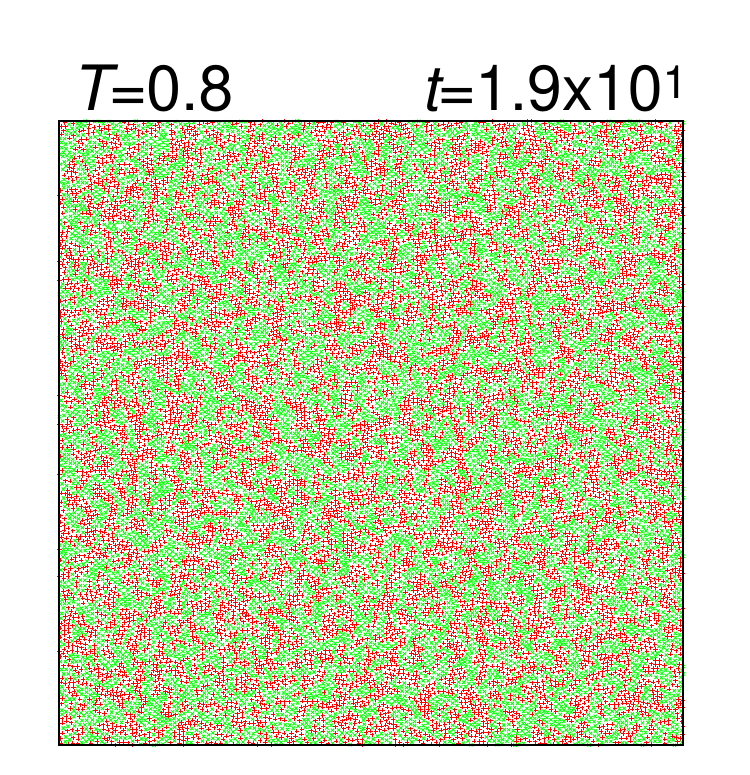}
\end{minipage}
\begin{minipage}{0.25\hsize}
\includegraphics[width=1.02\columnwidth]{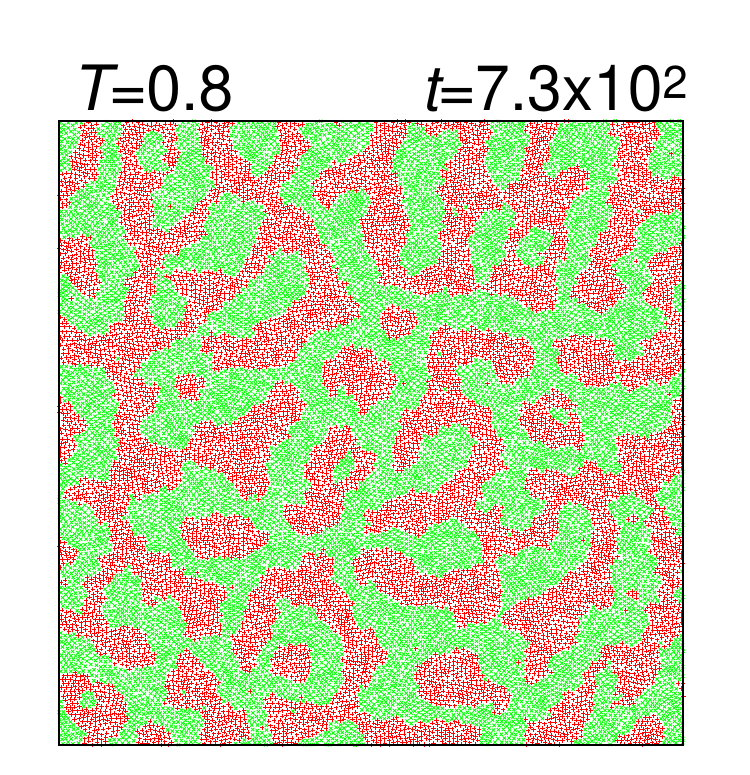}
\end{minipage}
\begin{minipage}{0.25\hsize}
\includegraphics[width=1.02\columnwidth]{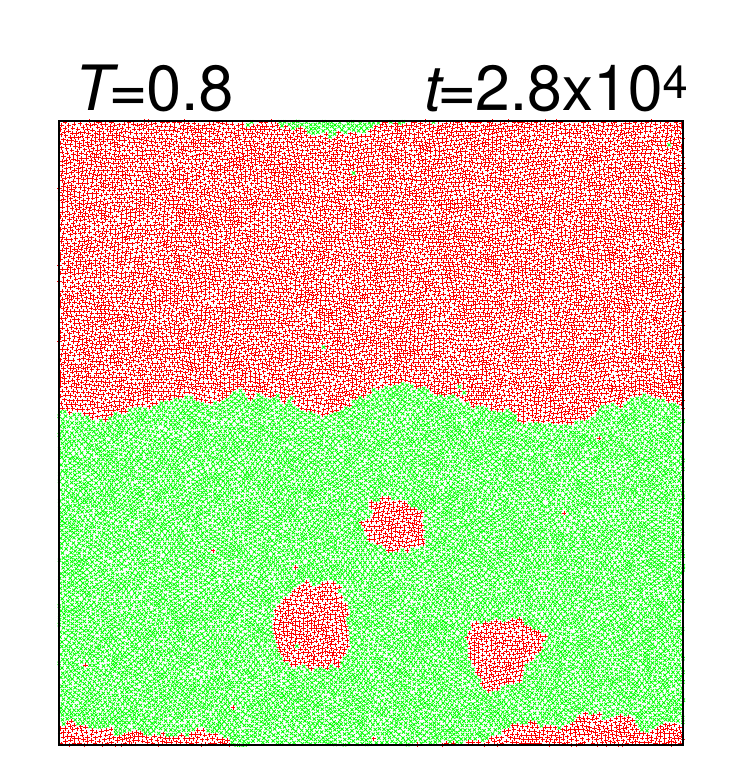}
\end{minipage}
\begin{minipage}{0.25\hsize}
\hspace{2cm}
\end{minipage}
\end{tabular}
\begin{tabular}{llll}
\begin{minipage}{0.25\hsize}
\includegraphics[width=1.02\columnwidth]{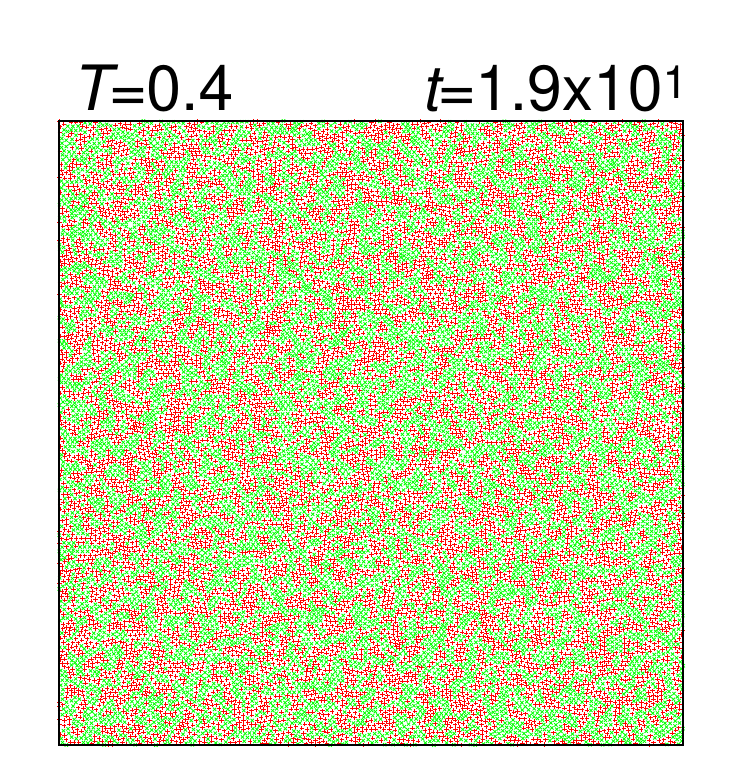}
\end{minipage}
\begin{minipage}{0.25\hsize}
\includegraphics[width=1.02\columnwidth]{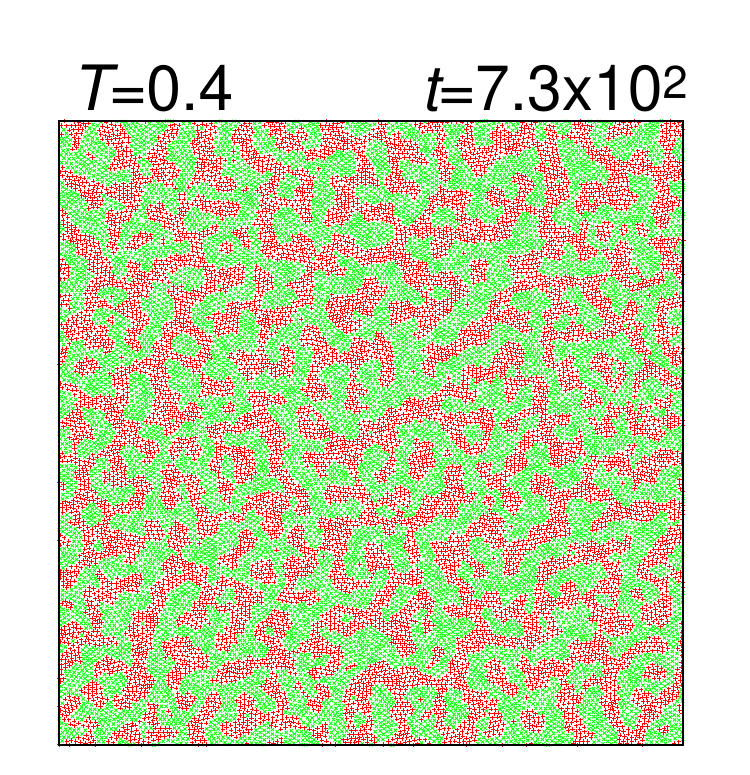}
\end{minipage}
\begin{minipage}{0.25\hsize}
\includegraphics[width=1.02\columnwidth]{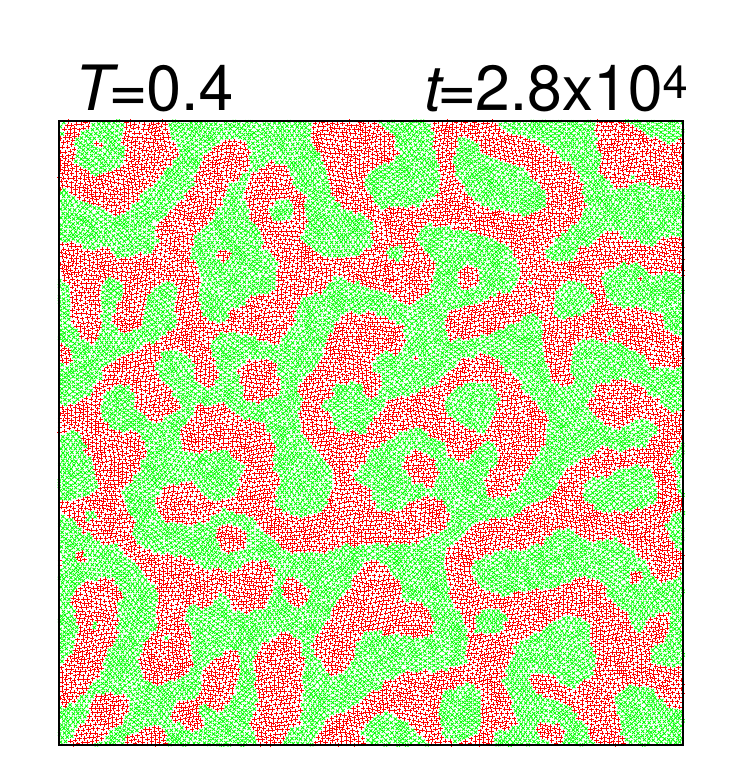}
\end{minipage}
\begin{minipage}{0.25\hsize}
\includegraphics[width=1.02\columnwidth]{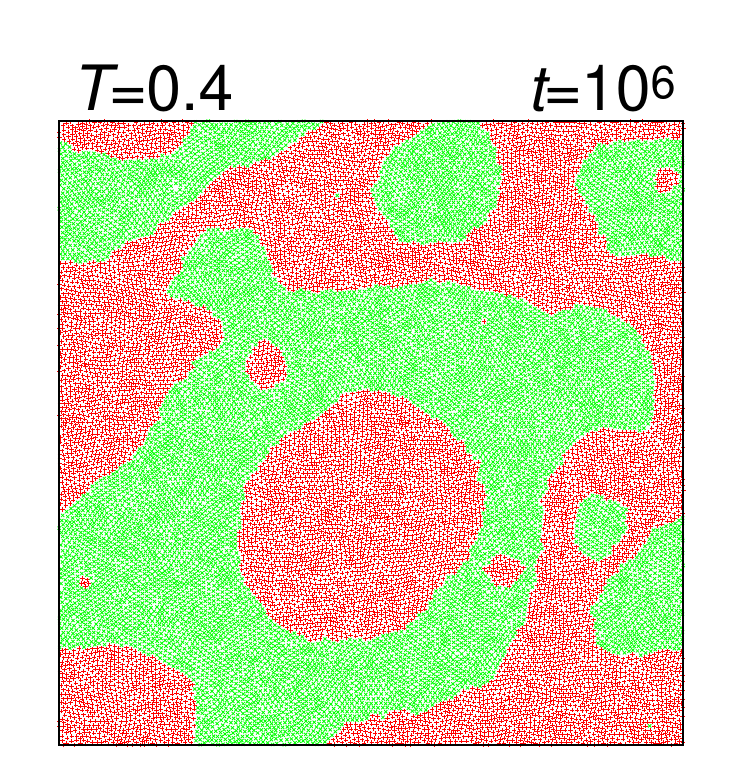}
\end{minipage}
\end{tabular}
\begin{tabular}{llll}
\begin{minipage}{0.25\hsize}
\includegraphics[width=1.02\columnwidth]{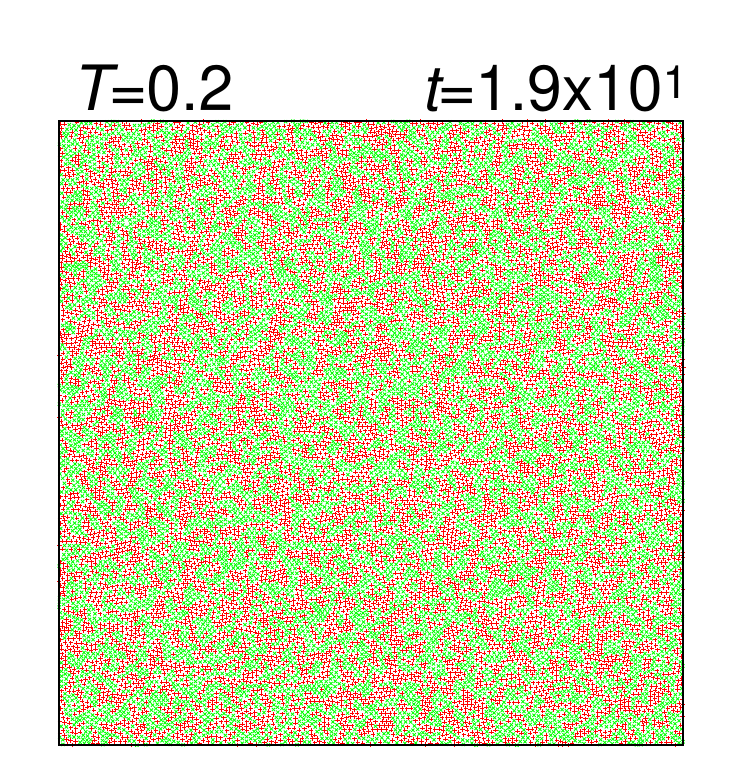}
\end{minipage}
\begin{minipage}{0.25\hsize}
\includegraphics[width=1.02\columnwidth]{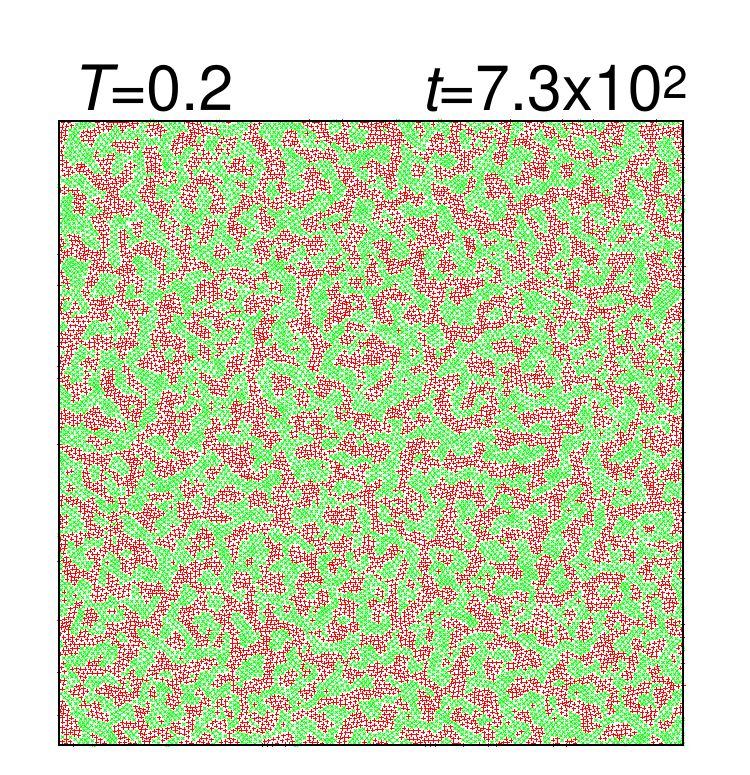}
\end{minipage}
\begin{minipage}{0.25\hsize}
\includegraphics[width=1.02\columnwidth]{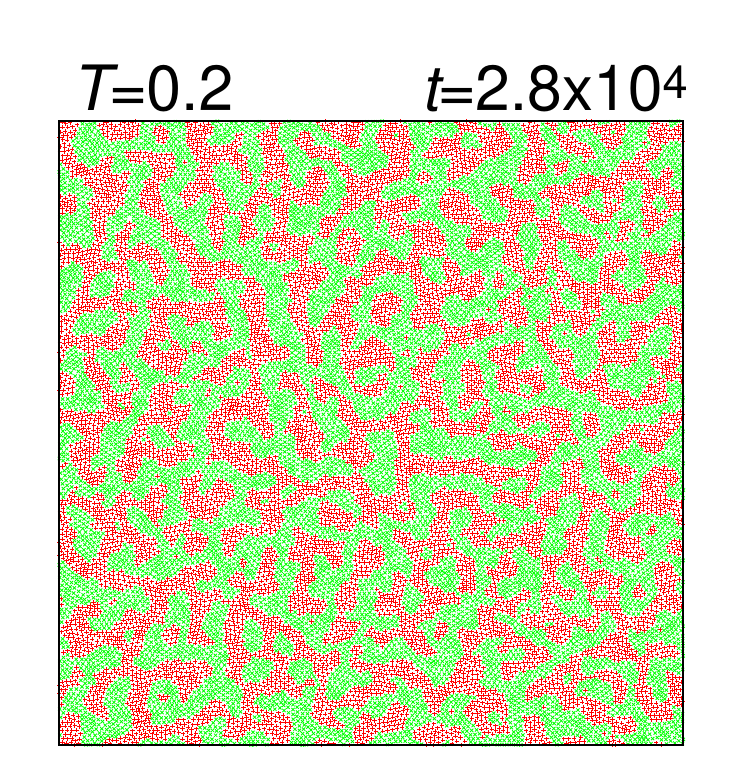}
\end{minipage}
\begin{minipage}{0.25\hsize}
\includegraphics[width=1.02\columnwidth]{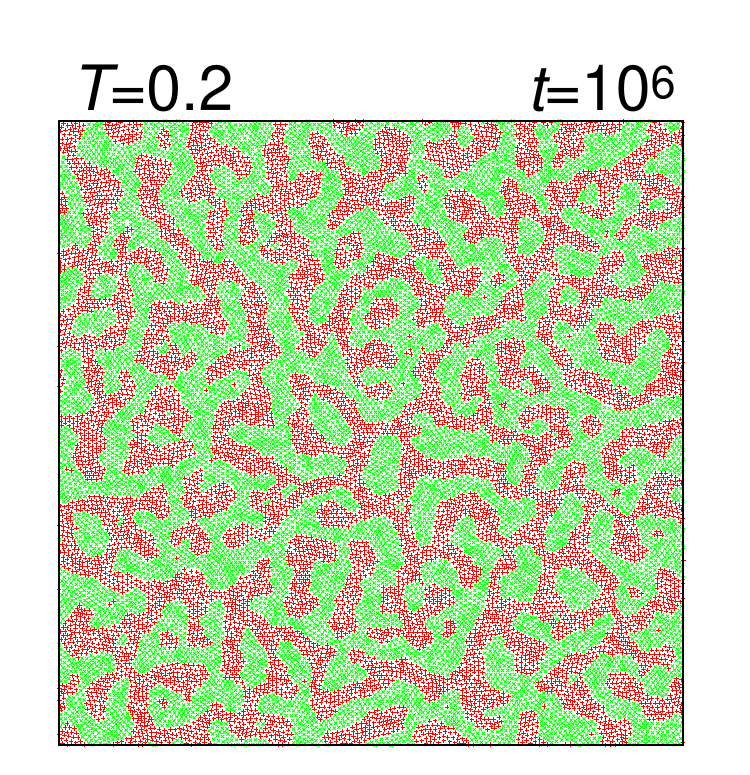}
\end{minipage}
\end{tabular}
\caption{
Snapshots obtained from the MD simulations at $T=0.8$ (top), 0.4 (middle), and 0.2 (bottom) at several times $t$.
The A and B particles are presented as red and green crosses, respectively.
}
\label{fig4}
\end{figure*}

We now study the phase separation kinetics of the binary model ($x=0.5$). 
We first equilibrate the system at $T=100$, suddenly quench it to the target temperatures ($T \leq 0.8$), and perform constant-temperature MD or MC simulations at each temperature.
For the simulation details, see Sec.~II~B and C.  
The equilibration and target temperatures are far above and below $T_c$, respectively; hence, the impact of the critical fluctuation near $T_c$ is negligible. 
We describe the results of the MD simulations in subsections A, B, and C and those of the MC simulations in D. 

In Fig.~\ref{fig4}, we show snapshots of the particles at $T=$ 0.8, 0.4, and 0.2 at several times $t$. 
$T=0.8$ is above $T_{\rm onset}$ and $T=0.2$ is below $T_{\rm g,sim}$ of the pure model. 
At the earliest time $t=1.9 \times 10^1$, the snapshots at all temperatures look almost the same; the A and B particles mix well. 
However, the snapshots at later times strongly depend on the temperature. 
At $T=0.8$, clear domains with domain walls already appear at $t=7.3 \times 10^2$, and the characteristic domain size becomes comparable to the system size at $t=2.8 \times 10^4$. 
At $T=0.4$, the domain size becomes comparable to the system size only at $t=10^6$. 
At $T=0.2$, the domain size does not reach the system size even at $t=10^6$. 
Therefore, the domain growth becomes drastically slower with decreasing temperature. 
However, interestingly, the snapshots at $t=2.8 \times 10^4$ and $10^6$ at $T=0.2$ are clearly different, suggesting that the phase separation proceeds very slowly even at $T < T_{\rm g,sim}$. 
In the next subsections, we analyze the time evolution of the domain size quantitatively. 

Although the kinetics of the domain growth strongly depend on the temperature, the morphology of the domains at various temperatures seem to be very similar. 
For example, the snapshot at $t=2.8 \times 10^4$ for $T=0.4$ looks similar to that at $t=7.3 \times 10^2$ for $T=0.8$. 
Similarly, the snapshot at $t=2.8 \times 10^4$ for $T=0.2$ looks similar to that at $t=7.3 \times 10^2$ for $T=0.4$. 
This observation suggests that the domain structures are statistically similar at all temperatures, though their kinetics are widely different. 
This point is studied in subsection C by analyzing the dynamic scaling of the correlation functions. 

\subsection{Characteristic domain size}

To measure the characteristic size of the domains, we calculate the chord length of the coarse-grained density field~\cite{Testard2014}. 
We first define the density field $\rho_l$ as
\begin{equation}
\rho_{l}(\vec{r},t) = \frac{1}{\pi l^2}\sum_{i \in {\rm A}} \theta(l - |\vec{r}-\vec{r}_i(t)|), 
\end{equation}
which is the mean density of A particles in a circle with center $\vec{r}$ and radius $l$. 
Then, using $\rho_l$, we define the coarse-grained density field as 
\begin{eqnarray}
\overline{\rho_l} (\vec{r},t) &=& \frac{1}{6}[2\rho_l (\vec{r},t) + \rho_l (\vec{r}+l\vec{e}_x,t) + \rho_l (\vec{r}-l\vec{e}_x,t) \nonumber \\
&& +\rho_l (\vec{r}+l\vec{e}_y,t) + \rho_l (\vec{r}-l\vec{e}_y,t)], 
\end{eqnarray}
where $\vec{e}_{\alpha}$ is the unit vector in the direction $\alpha$.
In practice, we set the coarse-grained length $l =1$, introduce equally spaced spatial grids with the distance $l$ in the simulation cell ($147 \times 147$ for $N=20000$), and calculate $\overline{\rho_l}$ at each grid point. 
Because the mean density of A particles in the simulation cell is $0.4625$, we regard the grid points with $\overline{\rho_l} > 0.4625$ as A-rich; otherwise, they are B-rich. 
We measure the chord lengths of the A-rich and B-rich regions along all vertical and horizontal grid lines and calculate the probability distribution of the chord length at time $t$. 
Then, we evaluate the characteristic domain size $\xi(t)$ as the first moment of the chord length distribution. 
Finally, we average $\xi(t)$ over the independent runs at the same $t$. 

\begin{figure}[t]
\centering
\includegraphics[width=\columnwidth]{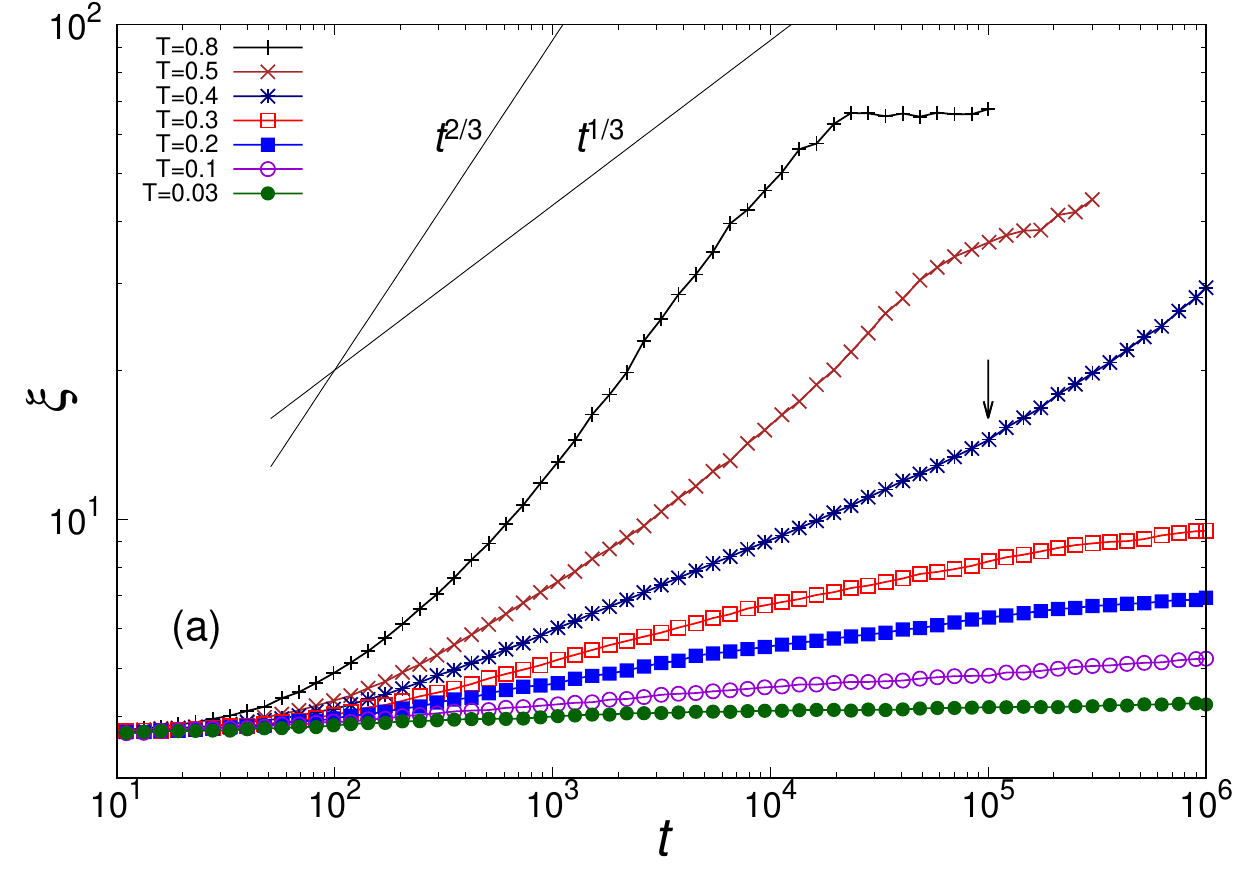}
\includegraphics[width=\columnwidth]{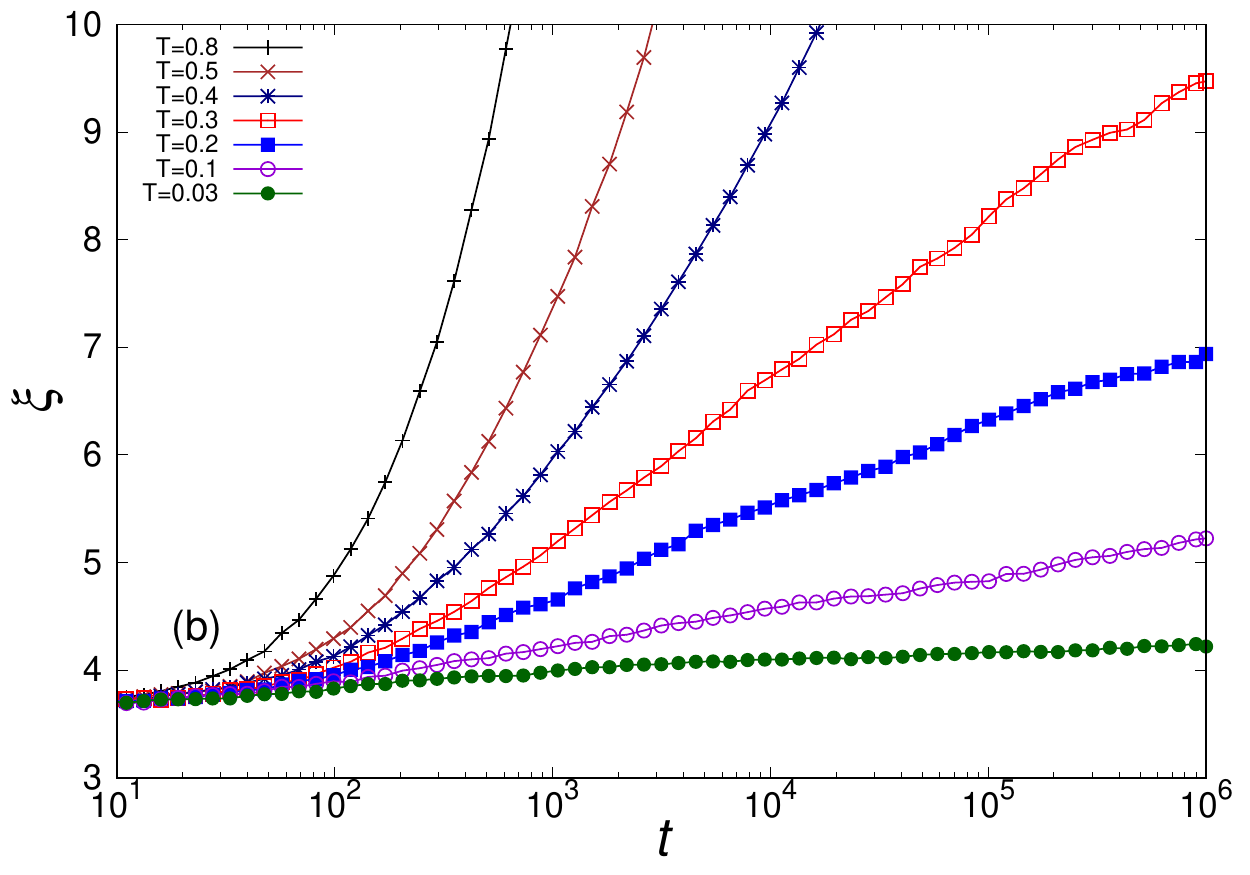}
\caption{
Domain size $\xi$ against time $t$ in the MD simulations. 
$\xi$ was estimated by the chord length method. 
Double logarithmic plot (a) and semi-logarithmic plot (b).  
The thin solid lines indicate the hydrodynamic coarsening law $\xi \propto t^{2/3}$ and diffusive coarsening law $\xi \propto t^{1/3}$. 
The arrow indicates $t=10^5$, i.e., the time scale of the crossover from a slower domain growth $\xi \propto t^{0.19}$ to the diffusive coarsening at $T=0.4$. 
}
\label{fig5}
\end{figure}

Fig.~\ref{fig5}(a) shows the double logarithmic plot of the domain size against time at various target temperatures. 
At the highest temperature $T=0.8$, we observe the fastest domain growth. 
The result is consistent with hydrodynamic coarsening $\xi \propto t^{2/3}$ in the long time region $t \gtrsim 10^3$ and saturates at $t \gtrsim 10^4$ due to the finite size effect. 
We note that the power law fitting of our data in $10^3 \leq t \leq 10^4$ gives the exponent 0.59, which is somewhat smaller than $2/3$. 
This small discrepancy is due to the velocity rescaling for the temperature control. 
In Ref.~\cite{Velasco1993}, the exponent 0.59 is reported for the velocity rescaling method, and 0.65, for the Nos\'{e}-Hoover thermostat. 
With decreasing temperature, the domain growth becomes noticeably slower. 
At $T=0.5 = T_{\rm onset}$, $\xi (t)$ is consistent with the diffusive coarsening $\xi \propto t^{1/3}$ at $t \gtrsim 10^3$. 
The growth becomes slightly faster at $t \gtrsim 10^4$, indicating a crossover from diffusive to hydrodynamic coarsening. 
However, we do not observe the clear hydrodynamic coarsening law at this temperature in our simulation, presumably due to the finite-size effect. 
At $T=0.4$, the domain growth becomes even slower. 
At $10^3 \leq t \leq 10^5$, our result is comparable to a weaker power law $\xi \propto t^{0.19}$. 
We note that we can observe clear domains with domain walls even in this weaker power law regime; the snapshot at $t=2.8 \times 10^4$ for $T=0.4$ in Fig.~\ref{fig4} corresponds to this regime. 
In the very long time region $t \gtrsim 10^5$ (indicated by an arrow), the domain growth becomes faster and consistent with diffusive coarsening $\xi \propto t^{1/3}$. 
At $T \leq 0.3 = T_{\rm g,sim}$, the domain growth is slower than any power law behaviors. 
To demonstrate this, we show the semi-logarithmic plot of $\xi(t)$ in Fig.~\ref{fig5}(b). 
$\xi (t)$ is nearly linear at $T=0.3$ and even concave at $T=0.2$ and 0.1. 
These results mean that the domain growth becomes logarithmically slow or even slower than logarithmic at these temperatures; however, we emphasize that it did not stop completely. 

In summary, we find that the domain growth does not stop but becomes drastically slow in the glassy regime. 
The onsets of hydrodynamic and diffusive coarsening are drastically delayed with decreasing temperature, and a weaker power law and further slower growth appear. 

\subsection{Structure factor}

\begin{figure}
\centering
\includegraphics[width=\columnwidth]{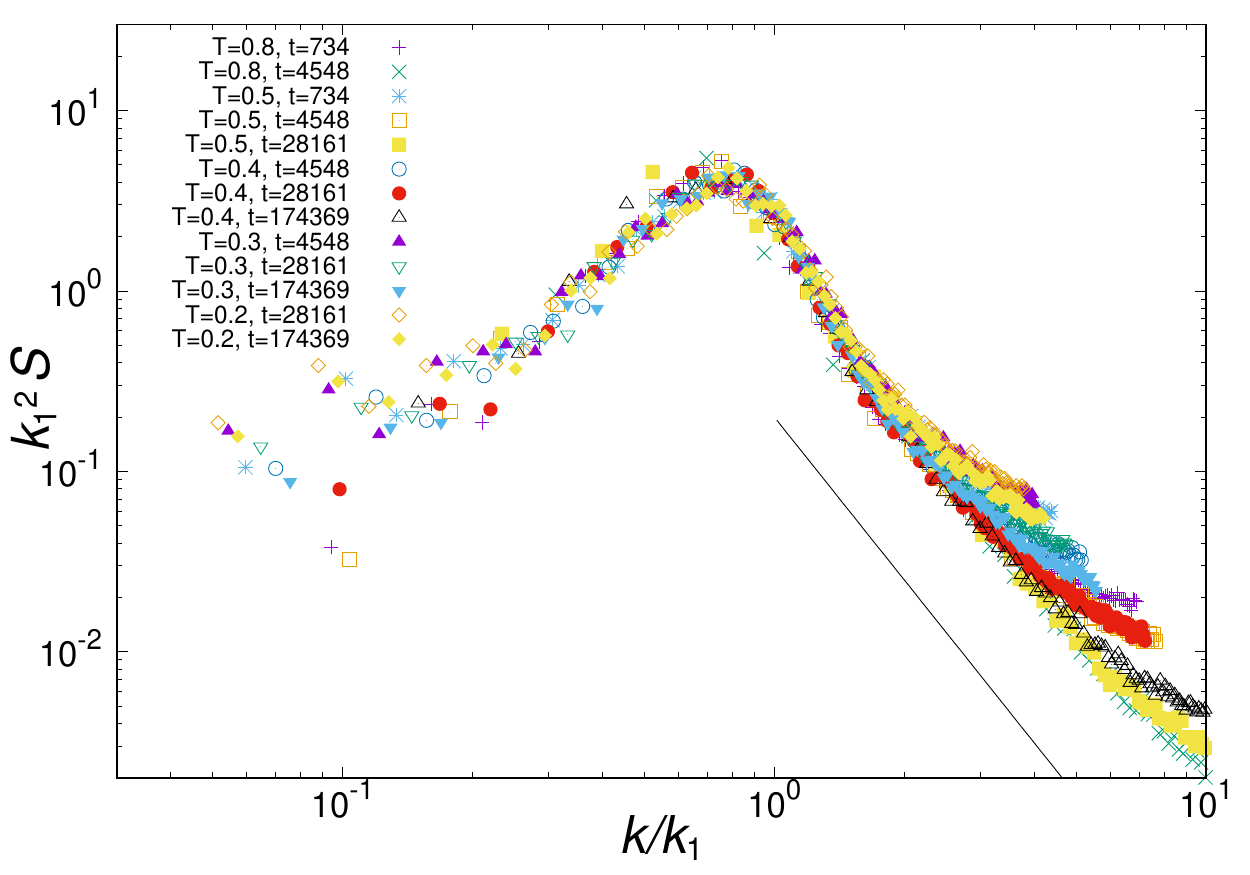}
\caption{
Dynamic scaling of the structure factor $S(k,t)$ at various temperatures and different times.
The thin solid line indicates Porod's law $S(k,t) \propto k^{-3}$.
}
\label{fig6}
\end{figure}

Another way to characterize the structure of the domains is to measure the spatial correlation functions, such as the structure factors. 
We define the structure factor of A particles at time $t$ as
\begin{equation}
S(k,t) = \frac{1}{N} \left| \sum_{i \in {\rm A}} e^{i \vec{k} \cdot \vec{r}_i(t)} \right|^2.  
\end{equation}
We calculate this function and average over the results for independent runs at the same $T$. 
Then, we calculate the first moment of the structure factor
\begin{equation}
k_1(t) = \frac{\int^{k_{\rm cut}}_0 k S(k,t)dk}{\int^{\rm k_{cut}}_0 S(k,t)dk}  
\end{equation}
and estimate the domain size as $\xi'(t) = 2\pi /k_1(t)$.
Here, we introduce the cutoff wave number $k_{\rm cut} = \pi$ to focus only on the large-scale structure~\cite{Yamamoto1994}. 
We find that the growth of $\xi'(t)$ is qualitatively the same as that of $\xi(t)$. 
For completeness, we present $\xi'(t)$ in the Appendix. 

One of the key features of the phase separation kinetics is the emergence of scale invariance. 
This can best be illustrated by the dynamic scaling law of the structure factor: 
\begin{equation}
S(k,t) = (k_1)^{-d} f(k/k_1), 
\end{equation}
where $f(x)$ is the scaling function. 
This law suggests that the only relevant length scale is the mean size of the domains, and the domain structures at different times are statistically the same when the length is rescaled by the mean size of the domains. 
To elucidate if this law applies for the various coarsening processes that appear in our model, we plot $k_1^2 S$ against $k/k_1$ at several $T$ and $t$ in Fig.~\ref{fig6}. 
Here, the data for all the temperatures with the domain size $\xi \geq 6$ are included, except for $t \geq 10^4$ at $T=0.8$ and $t \geq 5 \times 10^4$ at $T=0.5$, where saturation of the domain size is observed. 
Noticeably, all the results collapse well. 
The collapse of the data between $T=0.8$ and 0.5 suggests that the scaling function $f(x)$ is almost the same for diffusive coarsening and hydrodynamic coarsening. 
The same observation was reported for the real space correlation function at $d=3$~\cite{Ahmad2012}. 
However, we also note that more detailed characterizations of the domain structure may reveal the difference between the two cases~\cite{Wagner1998}.  
Furthermore, these high temperature data collapse very well with the data for $T=0.4$, 0.3 and 0.2. 
Namely, the dynamic scaling law works well over a very wide range of temperatures including the glassy regime, and the scaling function $f(x)$ is universal within the accuracy of our data. 
This means that the domain structures appearing in the very slow coarsening in the glassy regime are almost the same as those without the glassy dynamics. 
This result is consistent with the visual inspection of Fig.~\ref{fig4}. 
Moreover, we find that the scaling function follows the power law $f(x) \propto x^{-3}$ at $x \gtrsim 2$. 
This means that Porod's law $S(k,t) \propto k^{-(d+1)}$ holds well for $k \gg k_1$ over a wide temperature range, which reflects the presence of sharp domain walls. 

\subsection{Domain growth in the MC simulations}

In this subsection, we study the phase separation kinetics using MC simulations. 
The purpose is to elucidate the impact of the microscopic dynamical rule on the various coarsening processes observed in our model. 
Because hydrodynamic transport is active in the MD simulations but not in the MC simulations, we can discuss the impact of hydrodynamic transport by comparing these two simulations. 

\begin{figure}
\centering
\includegraphics[width=\columnwidth]{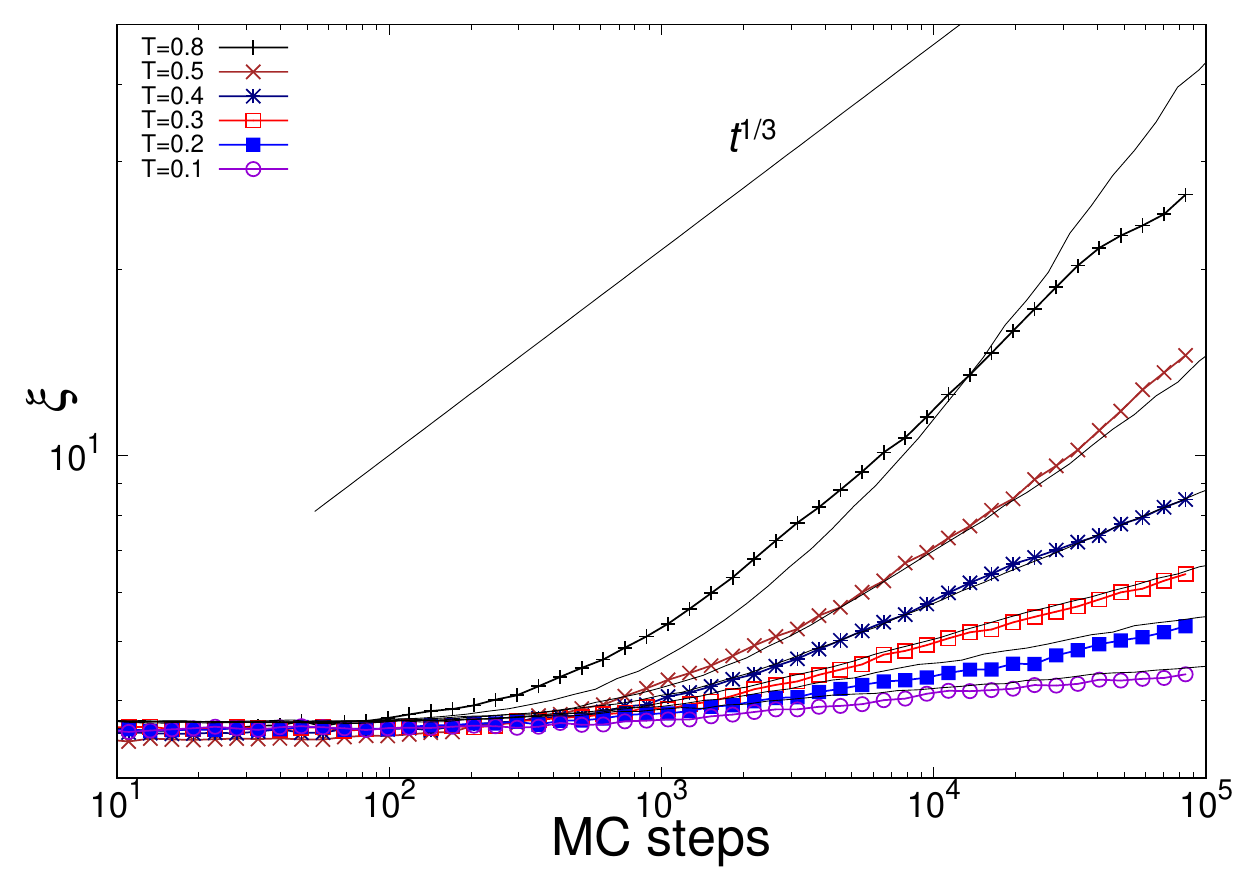}
\caption{
Domain size $\xi$ against the number of MC steps in the MC simulations. 
$\xi$ was estimated by the chord length method. 
The bottom six thin solid lines are the results of MD simulations at the corresponding temperatures, where the MD unit time is converted to 12 MC steps. 
The top thin solid line indicates the diffusive coarsening law $\xi \propto t^{1/3}$. 
}
\label{fig7}
\end{figure}

Fig.~\ref{fig7} shows the time evolution of the domain size (measured by the chord length method) in the MC simulations, compared with that in the MD simulations. 
To compare the MC and MD results on equal footing, we convert the time unit of MD to 12 MC steps, where 12 is chosen to maximize the overlap between two sets of data. 
At $T=0.8$, the MC and MD results are clearly different, and the MC simulation presents a slower domain growth. 
The MC result is reasonably consistent with the diffusive coarsening law $\xi \propto t^{1/3}$. 
This outcome confirms that the phase separation in the MD simulations is accelerated by the hydrodynamic transport at $T=0.8$. 
On the other hand, at $T \leq 0.5 = T_{\rm onset}$, the MC and MD results are quantitatively the same within our MC simulation time. 
Therefore, the very slow coarsening in the glassy regime is independent of the choice of MD or MC dynamics, meaning that the hydrodynamic effect is negligible and the diffusion of particles is the dominant mechanism of the phase separation in this regime. 
The irrelevance of the microscopic dynamical rule is one of the features of glassy dynamics~\cite{Berthier2011b}. 
Our results establish that this irrelevance also holds in the domain growth in the glassy regime. 

\section{Impact of the slowing down of the microscopic dynamics}

\subsection{Numerical results}

\begin{figure*}[t]
\begin{tabular}{lll}
\begin{minipage}{0.33\hsize}
\includegraphics[width=\columnwidth]{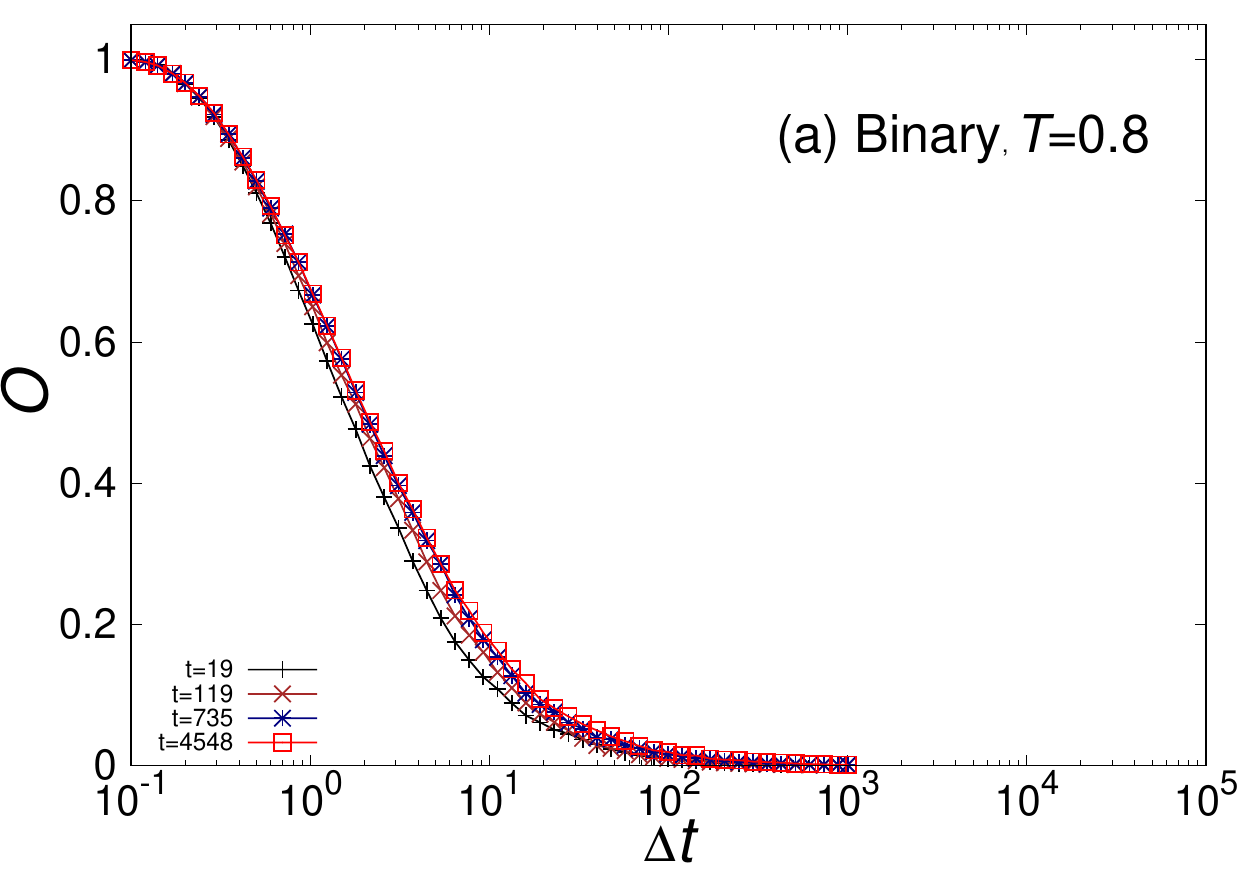}
\end{minipage}
\begin{minipage}{0.33\hsize}
\includegraphics[width=\columnwidth]{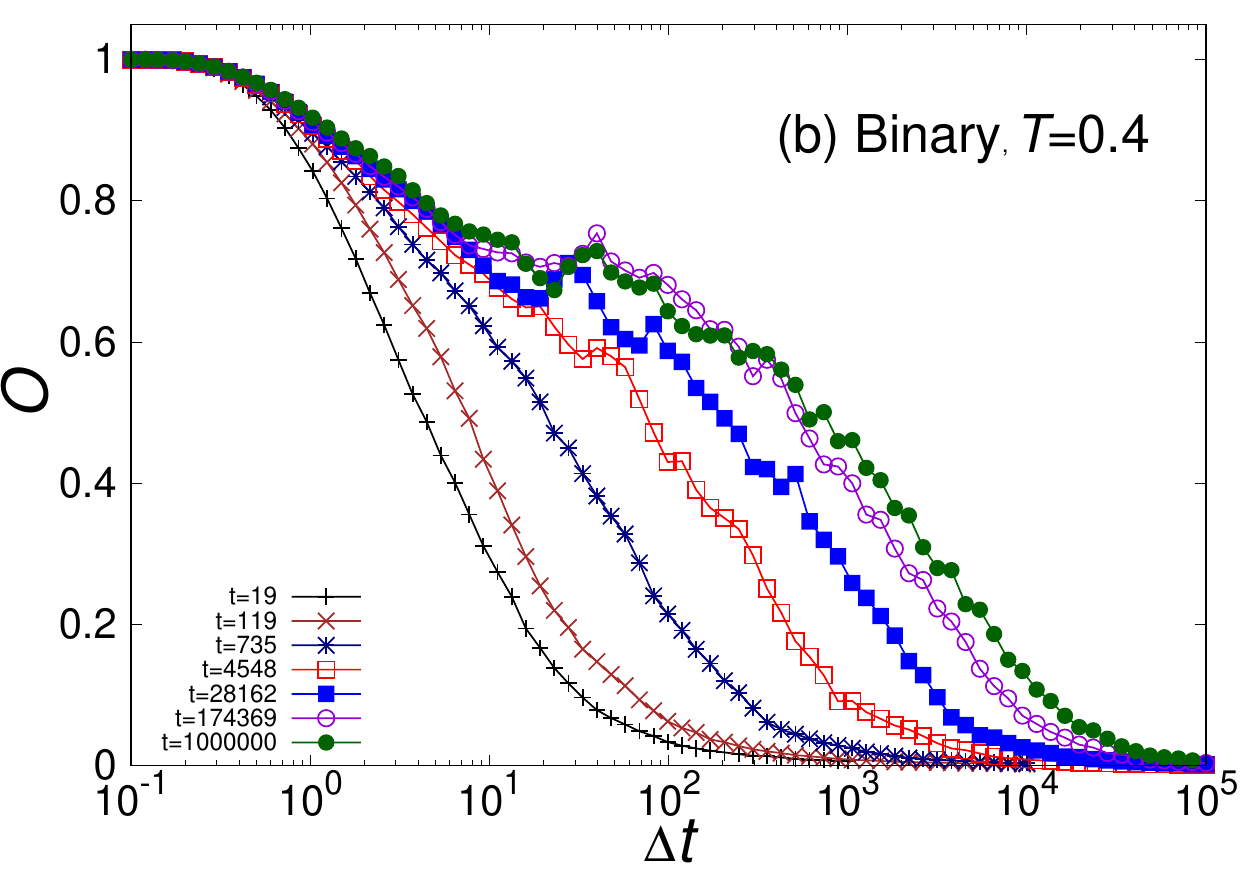}
\end{minipage}
\begin{minipage}{0.33\hsize}
\includegraphics[width=\columnwidth]{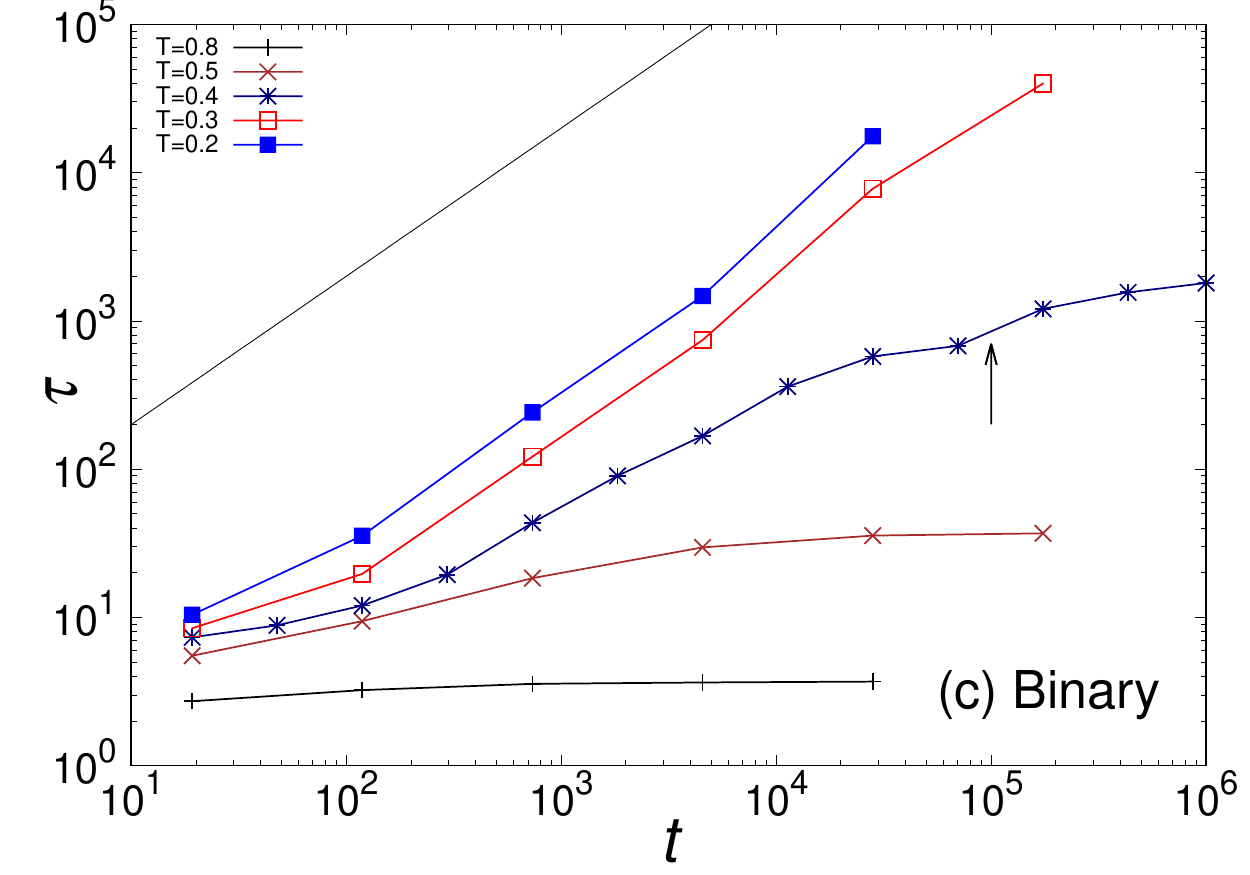}
\end{minipage}\end{tabular}
\begin{tabular}{lll}
\begin{minipage}{0.33\hsize}
\includegraphics[width=\columnwidth]{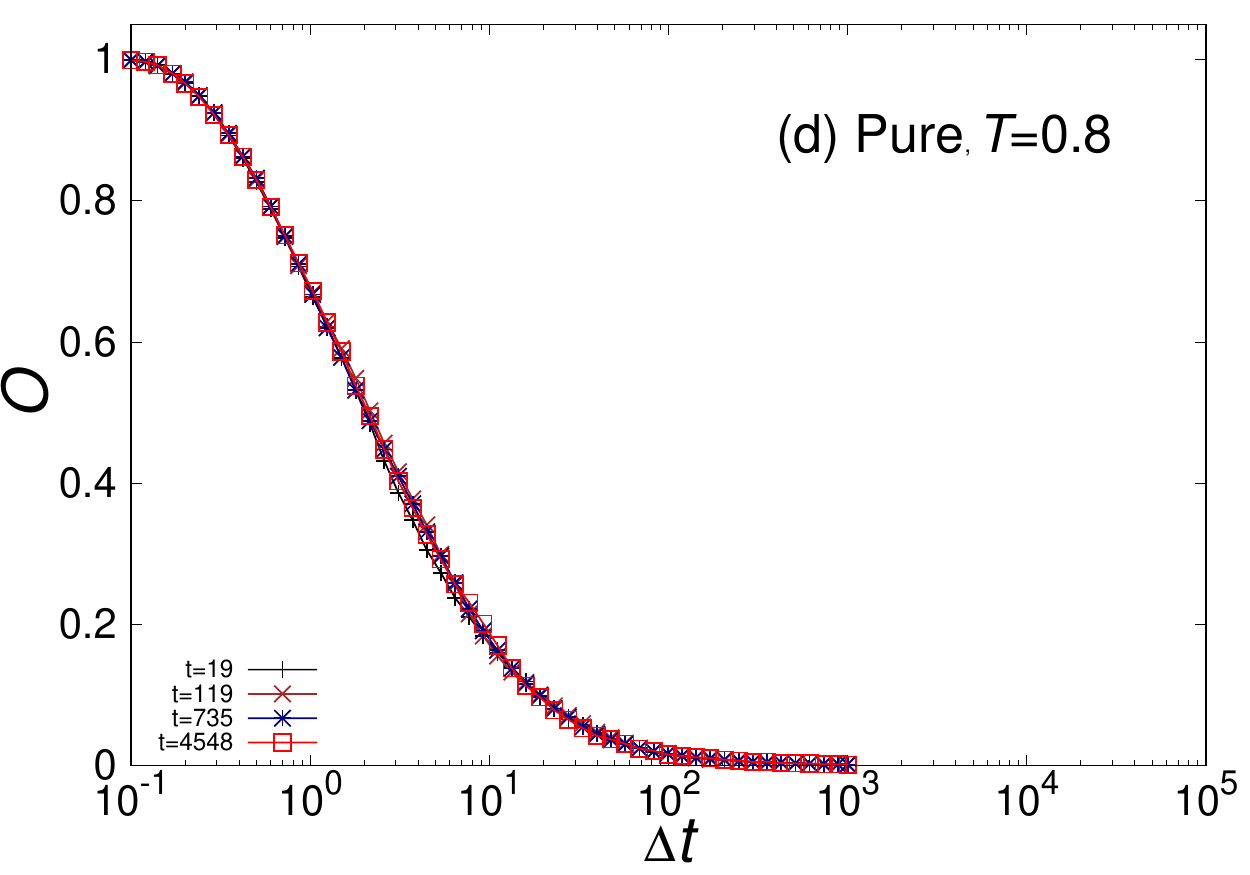}
\end{minipage}
\begin{minipage}{0.33\hsize}
\includegraphics[width=\columnwidth]{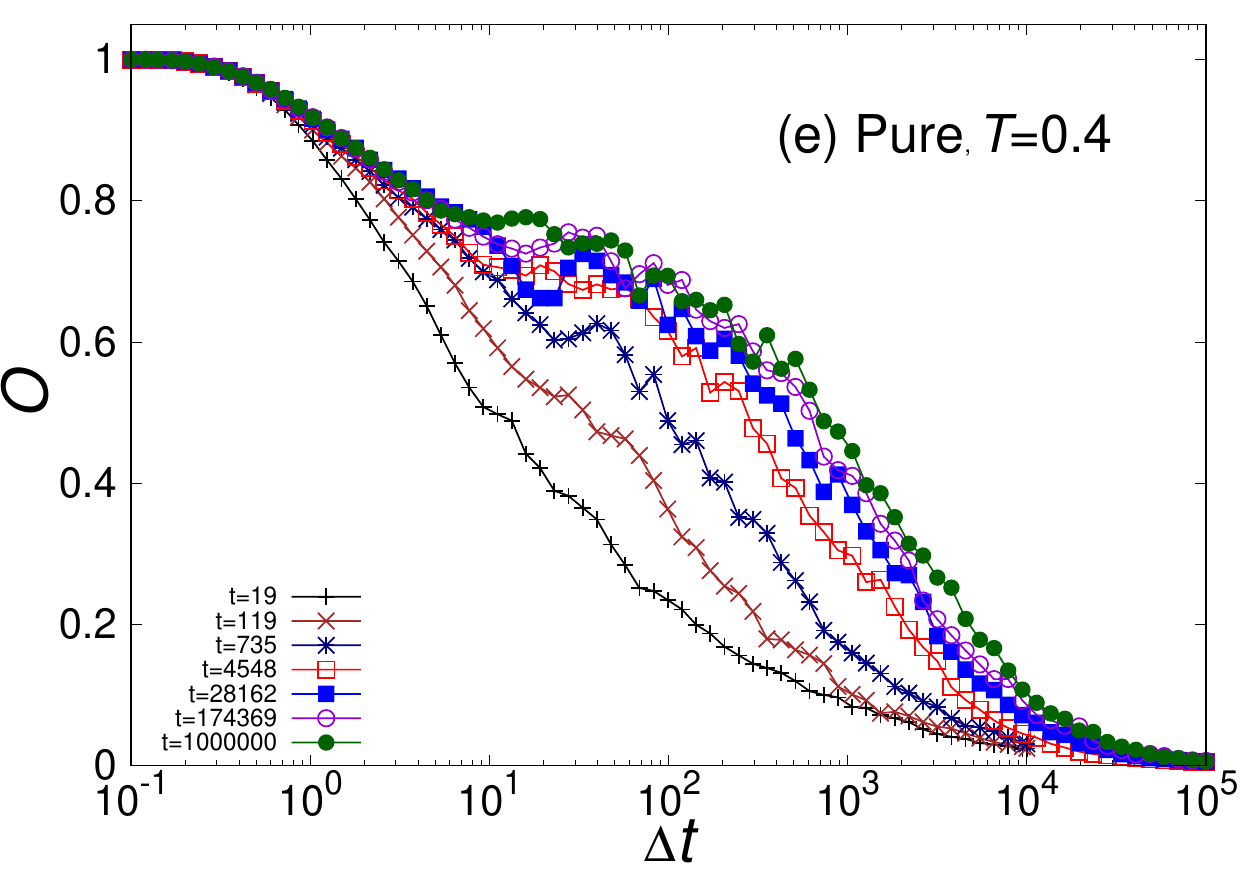}
\end{minipage}
\begin{minipage}{0.33\hsize}
\includegraphics[width=\columnwidth]{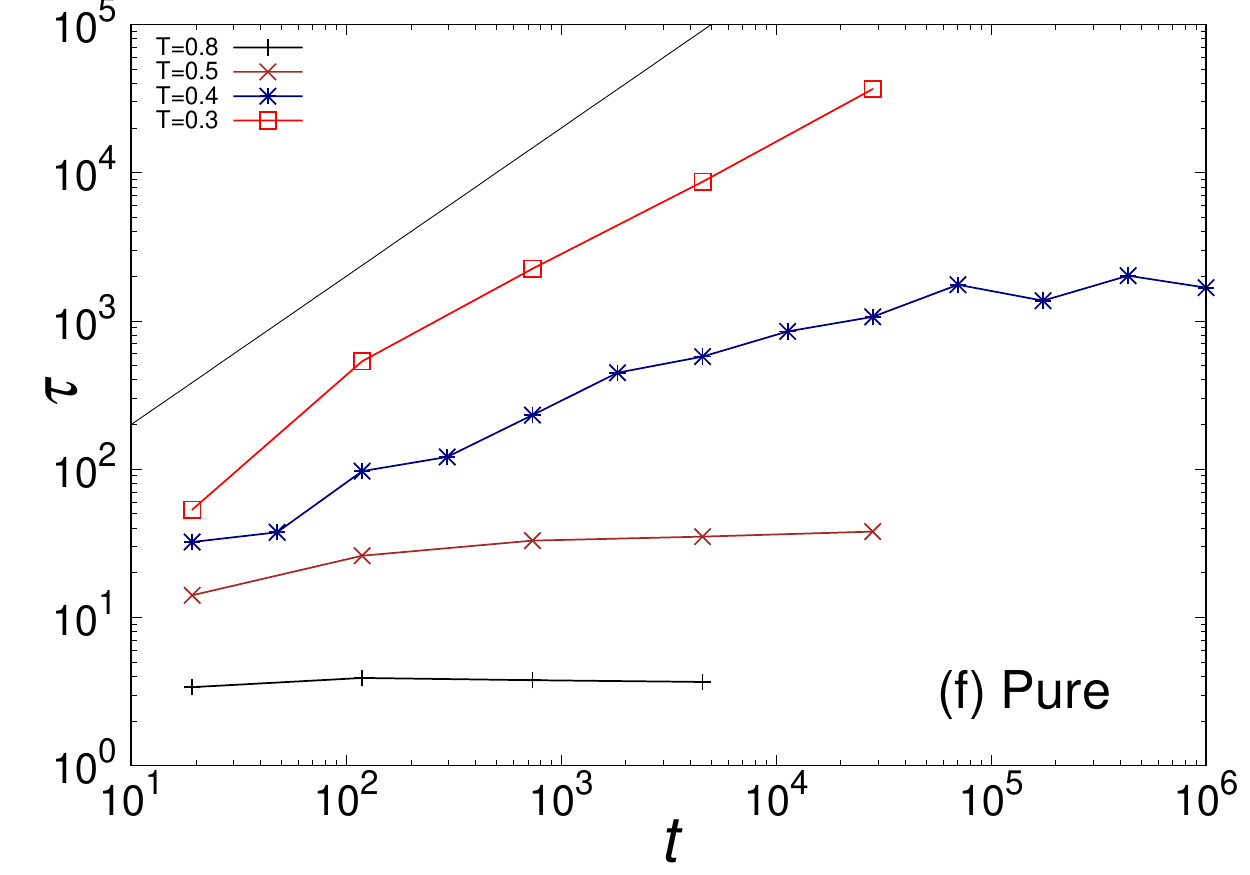}
\end{minipage}
\end{tabular}
\caption{
Waiting-time dependence of the self-part of the overlap function $O(\Delta t, t)$ and the relaxation time $\tau(t)$. 
Binary model during phase separation (a,b,c), and pure model without phase separation (d,e,f). 
The overlap functions are measured for the displacements of particles between time $t$ and $t + \Delta t$ at each temperature.  
The solid lines in (c,f) indicate $\tau \propto t$. 
The arrow in (c) indicates $t = 10^5$, at which the diffusive coarsening sets in at $T=0.4$ (see Fig.~\ref{fig5}). 
}
\label{fig8}
\end{figure*}

We find that the slow domain growth, which is much slower than diffusive coarsening, appears in the low-temperature region of $T \leq 0.4$. 
To gain insight into this coarsening process, we here analyze the microscopic relaxation dynamics during the phase separation. 
We calculate the two-time version of the overlap function: 
\begin{eqnarray}
O(\Delta t, t) = \frac{1}{N} \sum_{i} \theta (|\vec{r}_i(t + \Delta t) - \vec{r}_i(t)| - \ell),
\end{eqnarray}
with $\ell=0.3$. 
This function quantifies the microscopic relaxation of particles between time $t$ and $t + \Delta t$. 
We emphasize that this function can decay without any growth of the domains just by, e.g., the exchange of particles of the same species. 
We calculate this function for independent runs and take the average over them at the same $T$. 
We also calculate the $t$-dependent relaxation time $\tau(t)$ as $O(\Delta t = \tau(t), t) = e^{-1}$. 
Note that the same function has frequently been studied in the context of the aging dynamics of glasses (without phase separation), in which case $t$ was called the waiting time. 
In this case, the experiments revealed the power law growth of the relaxation time $\tau(t) \propto t^{\alpha}$ with the aging exponent $\alpha \in [0.5,1]$~\cite{Berthier2011b,struik1978physical,Kob1997}. 
Theoretically, a trap model exhibits $\alpha = 1$ in the asymptotically long-waiting-time regime~\cite{Berthier2011b,Monthus1996}. 

Fig.~\ref{fig8}(a,b) show $O(\Delta t, t)$ of the binary model during the phase separation. 
At $T=0.8$, the overlap functions quickly decay to zero at all waiting time $t$. 
The decay becomes slightly slower with $t$, and it becomes completely independent of $t$ at $t \geq 735$. 
This slowing is very weak; the relaxation time at $t \geq 735$ is only 1.5 times larger than that at $t = 19$.
In contrast, at $T=0.4$, the decay strongly depends on the waiting time. 
At $t=19$, the overlap function decays quickly as in the case at higher temperatures. 
However, the decay becomes increasingly slower with $t$, and the relaxation time at $t=10^6$ becomes alomst $10^3$ times larger than that at $t=19$. 
We also find that the overlap functions at the two largest $t$ are almost the same, suggesting that they finally become independent of the waiting time in the long-waiting-time region. 

Fig.~\ref{fig8}(c) summarizes the relaxation time $\tau(t)$ at various temperatures $T$. 
At $T=0.8$, the relaxation time $\tau$ is nearly independent of the waiting time $t$. 
The increase in $\tau$ with $t$ becomes more and more drastic with decreasing temperature. 
At $T \leq 0.3$, the relaxation time increases with $t$ without limitation in our simulation time. 
This slowing down is close to $\tau \propto t$ (solid line), which is very similar to the aging of glasses with an aging exponent $\alpha \approx 1$. 
At the intermediate temperature $T=0.4$, the relaxation times increase with $t$ but finally saturate in the long-waiting-time region. 
The slowing down at this temperature is milder than that at lower temperatures. 
By fitting the data at $10^3 \leq t \leq 10^5$, we obtain the aging exponent $\alpha \approx 0.6$. 
These microscopic slowing down are very similar to the aging dynamics of glassy systems without phase separation. 

We now compare these aging-like microscopic slowing down (Fig.~\ref{fig8}) with the kinetics of the domain growth (Fig.~\ref{fig5}).
At $T=0.8$ and 0.5, hydrodynamic and diffusive coarsening processes appear at $t \geq 10^3$, respectively. 
At these temperatures, the relaxation time depends only weakly on the waiting time in the same waiting-time region. 
At $T = 0.4$, the slow coarsening $\xi \propto t^{0.19}$ takes place at small $t$, and diffusive coarsening sets in at large $t$. 
For the microscopic dynamics, the relaxation time increases at small $t$ and saturates at large $t$. 
The crossover in the domain growth kinetics takes place at $t \approx 10^5$. 
Fig.~\ref{fig8} clearly shows that the crossover in the relaxation time also takes place at a similar time scale, as indicated by the arrow. 
At $T \leq 0.3$, the domain growth is logarithmically slow or even slower, and the relaxation time increases with $t$ without limitation. 
Therefore, all these results suggest that the domain growth becomes slower than diffusive coarsening when the microscopic dynamics exhibit the aging-like slowing down. 

Finally, we directly compare the microscopic slowing down during phase separation with the aging dynamics without phase separation. 
To this end, we calculate $O(\Delta t,t)$ of the pure model. 
We first equilibrate the pure model at $T=100$, suddenly quench the system, and then perform constant-temperature MD simulations at the target temperature. 
For the simulation details, see Sec.~II~B.
The obtained overlap functions and the relaxation time are plotted in Fig.~\ref{fig8}(d,e,f). 
We did not try to analyze $T \leq 0.2$ because the overlap function does not decay to zero even at small $t$ in this temperature region. 
The qualitative similarity between the results of the binary and pure models is clear. 
In both cases, the relaxation does not strongly depend on the waiting time at higher temperatures, but it does at lower temperatures. 
The relaxation time increases with the waiting time and finally saturates at $T = 0.4$, while it increases without limitation at $T = 0.3$.   
However, we also find quantitative differences between the two cases. 
The relaxation time in the binary model is shorter than that in the pure model when they are compared at the same waiting time. 
For example, at $T=0.4$, $\tau(735) = 43$ in the binary model, while $\tau(735) = 232$ in the pure model. 
Thus, we conclude that the microscopic slowing down during the phase separation in the glassy regime is similar to the aging dynamics without phase separation, but in the former case, the presence of the inhomogeneous concentration fields accelerates the microscopic relaxation. 

\subsection{Discussion}

The results in subsection A suggest that the slow domain growth in the glassy regime is accompanied by microscopic slowing down. 
This slowing down is similar to aging but is affected by the phase separation itself. 
For a faithful description of this situation, one has to consider the coupling between the phase separation kinetics and the aging dynamics in a glassy system. 

Instead of directly addressing this problem, we consider here a simple extension of the Cahn-Hilliard equation to discuss a possible link between the slow domain growth and the aging-like microscopic slowing down. 
The diffusive coarsening law is understood based on the Cahn-Hilliard equations~\cite{1992solids,Bray1994,onuki2007phase}: 
\begin{eqnarray}
\frac{\partial \phi}{\partial t} = L \nabla^2 \mu, \label{CH}
\end{eqnarray}
where $\phi$ is the concentration field of one component, $L$ is the local mobility coefficient, and $\mu$ is the chemical potential of the component~\footnote{For simplicity, we focused on the model in which $L$ does not depend on the spatial position.}. 
Typically, $L$ is assumed to be independent of time. 
This assumption is reasonable for high-temperature fluids because their microscopic dynamics do not strongly depend on the waiting time (see Fig.~\ref{fig8}(a)). 
However, at low temperatures, we observe a microscopic slowing down with the waiting time (see Fig.~\ref{fig8}(b)).
This observation suggests that $L$ is no longer a constant but decreases with time. 

Taking into account a possible time dependence of the mobility coefficient, we consider the continuum model described by Eq.~\ref{CH} with $L(t) = L_0 (1 + t/t_0)^{-\alpha}$, where $L_0$ is the mobility coefficient without glassy dynamics, $t_0$ is the microscopic time constant, and $\alpha$ is the aging exponent. 
To obtain the domain growth law in the late stage, we adopt the scaling analysis discussed in Refs.~\cite{Bray1994,onuki2007phase}. 
Note that we can obtain the same result by considering the time evolution of a spherical domain~\cite{1992solids}.
In the late stage, well-defined domains appear in the system, and the domain walls slowly move due to the chemical potential gradient. 
By focusing on this process, one can rewrite Eq.~\ref{CH} as $v = - L [\vec{n} \cdot \vec{\nabla} \mu]$, where $v$ is the velocity of the domain wall, $\vec{n}$ is the unit vector normal to the domain wall, and $[]$ denotes the discontinuity across the domain wall. 
Assuming that the only relevant length scale is the typical domain size $\xi$, the velocity of the domain wall can be estimated as $v \sim d \xi /d t$. 
Considering also the Gibbs-Thomson boundary condition, the gradient of the chemical potential can be estimated as $[\vec{n} \cdot \vec{\nabla} \mu] \sim \sigma {\cal K}/\xi \sim \sigma /\xi^2$, where $\sigma$ is the surface tension and $\cal K$ is the curvature of the domains. 
As a result, we obtain 
\begin{eqnarray}
\frac{d \xi}{d t} \sim \frac{L(t) \sigma}{\xi (t)^2}, 
\end{eqnarray}
the solution to which is 
\begin{eqnarray}
\xi(t)^3 \sim \sigma \int^{t}_{0} L(s) ds. \label{ans}
\end{eqnarray}
In the case of $\alpha < 1$, Eq.~\ref{ans} gives $\xi(t) \propto t^{(1 - \alpha)/3}$ in the long time region $t \gg t_0$. 
Therefore, in this model, the microscopic slowing down with $\alpha < 1$ leads to weaker power law domain growth with the exponent $(1-\alpha)/3$. 
In the case of $\alpha = 1$, Eq.~\ref{ans} gives $\xi(t) \propto (\log t )^{1/3}$ in the long time region $t \gg t_0$. 
The domain growth does not stop but becomes extremely slow. 

These results are not very unreasonable with respect to our MD simulation results. 
In our simulation at $T=0.4$, we observe microscopic slowing down with the aging exponent $\alpha \approx 0.6$ in $10^3 \leq t \leq 10^5$. 
In this time region, we observe a weaker power law growth of the domains. 
At $T \leq 0.3$, we find the aging exponent $\alpha \approx 1$ within all our simulation times. 
The domain growth in this case also does not stop but becomes logarithmically slow or even slower. 
Therefore, although very simplified, the continuum model qualitatively reproduces the slow domain growth in our numerical simulations. 
Quantitatively, however, the continuum model underestimates the rate of the domain growth. 
The aging exponent $\alpha \approx 0.6$ gives the domain growth exponent $(1-0.6)/3 \approx 0.13$, which is smaller than the observed value $0.19$. 
This discrepancy might be due to the omission of the heterogeneity in the glassy dynamics. 

We remark that during these periods of slow domain growth, the domains are separated by domain walls (Fig.~\ref{fig4}), and the dynamic scaling law of the structure factors holds (Fig.~\ref{fig6}).
This observation suggests that the slow coarsening process cannot be seen as the early stage of phase separation; therefore, it is reasonable to compare our MD results with those of the late-stage analysis.   

\section{Conclusions}

In this paper, we used MD simulations to study the simplest example of the glass-glass phase separation. 
The model consists of type A and B particles, where the A-A and B-B interactions have identical LJ potential and the A-B interaction is purely repulsive. 
To avoid crystallization, we also introduced polydispersity in the sizes of the particles. 
We mainly studied the 50:50 mixture (called the binary model) but also considered the 100:0 case (called the pure model) to gain insight into the microscopic dynamics. 
We first calculated the coexisting temperatures of A-rich and B-rich fluids using the semi-grand canonical MC simulations. 
Next, we studied the equilibrium dynamics of the pure model to determine the characteristic temperatures of the glassy dynamics. 
We showed that the onset temperature of the glassy dynamics $T_{\rm onset}$ and the simulation glass transition temperature $T_{\rm g,sim}$ are located deep inside the immiscible region in the phase diagram. 
Then, we studied the kinetics of the phase separation at various temperatures. 
At $T > T_{\rm onset}$, the domain growth was fast and consistent with hydrodynamic coarsening $\xi \propto t^{2/3}$. 
However, at $T_{\rm onset} > T > T_{\rm g,sim}$, we observed a weaker power law growth within a shorter time and crossover to diffusive coarsening $\xi \propto t^{1/3}$ within a longer time. 
At $T < T_{\rm g,sim}$, the domain growth became logarithmically slow or even slower than logarithmic within all our simulation times. 
Despite the drastic slowing down of the domain growth, we found that the structure factor at various temperatures followed the dynamic scaling law very well under the same scaling function, meaning that the domain structures in the glassy regime were statistically similar to those at higher temperatures. 
By comparing the MD results with the MC results, we established that this slow coarsening in the glassy regime was not affected by hydrodynamic transport. 
Finally, we analyzed the microscopic dynamics during the phase separation. 
The microscopic dynamic was almost independent of the waiting time at $T > T_{\rm onset}$ but strongly dependent on the waiting time at $T < T_{\rm onset}$. 
We found that the domain growth became slower than diffusive coarsening when the microscopic dynamics exhibited the drastic slowing down. 
This slowing down has similarities with the aging dynamics without phase separation. 
However, quantitatively, the relaxation of the binary model during the phase separation was faster than that of the pure model without phase separation, suggesting that the inhomogeneous concentration field accelerates the microscopic relaxation in the former case. 
We discussed a possible mechanism of the slow domain growth in the glassy regime by a simple extension of the Cahn-Hilliard equation to take into account the aging-like microscopic slowing down. 

In this work, we analyzed only the domain growth kinetics and microscopic relaxation dynamics. 
However, it is known that various physical properties, such as mechanical and electrical properties, change with the phase separation. 
This phenomenon is true for various solids~\cite{onuki2007phase,Kato2019} and for phase-separating glasses~\cite{Doremus1994}. 
It should be interesting to use an MD simulation to study these physical properties.


\begin{acknowledgments}
This work was supported by the Japan Society for the Promotion of Science (JSPS) Grants-in-Aid for Scientific Research (KAKENHI) Grants No. 16H04034, 17H04853, 17H06375, 18H05225, and 19H01812. 
\end{acknowledgments}

\appendix
\section{Domain size measured by structure factors}
\begin{figure}[h]
\centering
\includegraphics[width=\columnwidth]{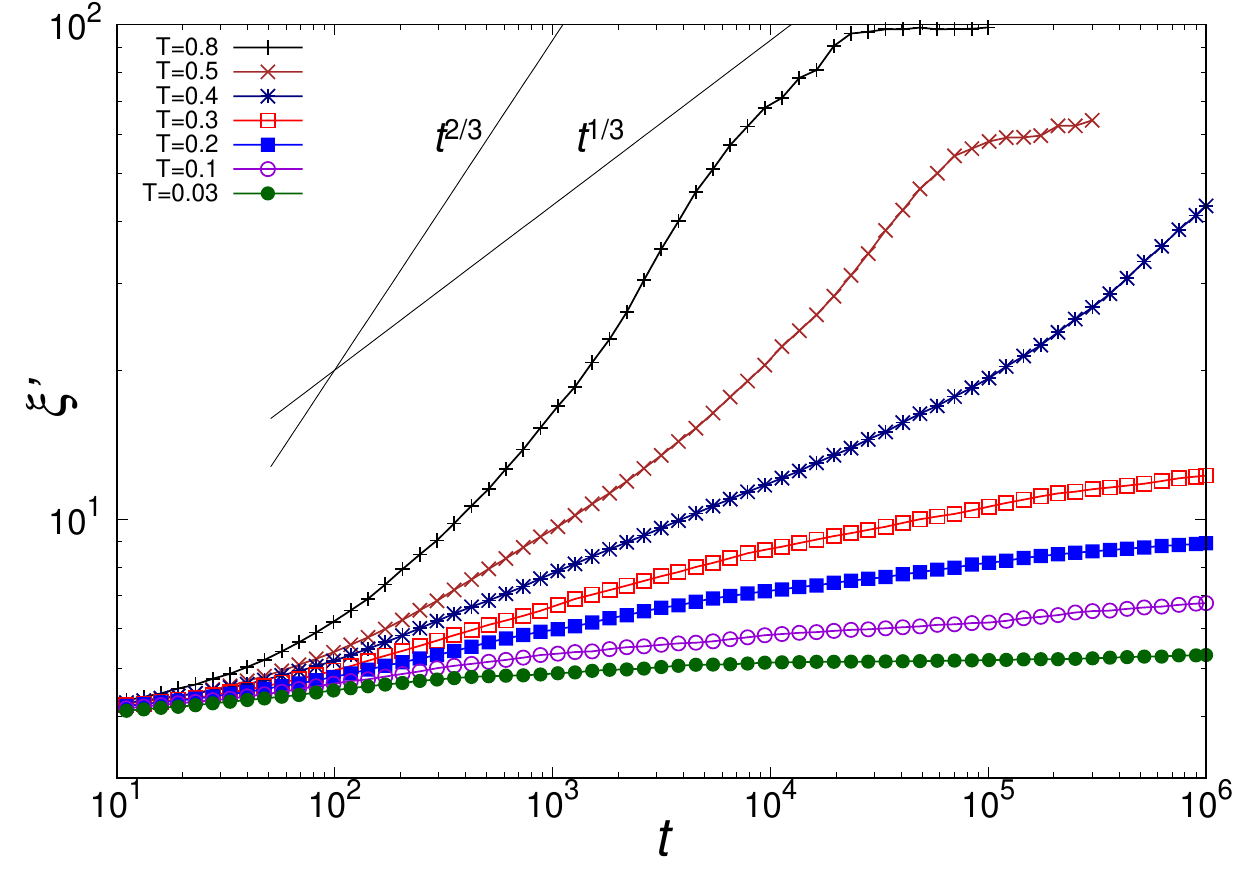}
\caption{
Domain size $\xi'$ against time $t$ in the MD simulations. 
$\xi'$ was estimated by the first moment of the structure factor. 
}
\label{figa}
\end{figure}

Fig.~\ref{figa} presents the time evolution of the domain size $\xi'$ measured by the first moment of the structure factor. 
The results are qualitatively the same as those obtained by the chord length method in Fig.~\ref{fig5}.

\bibliographystyle{apsrev4-1}
\bibliography{final_bib}

\end{document}